\documentclass[12pt,preprint]{aastex}
\usepackage{emulateapj5,epsf}

\makeatletter
\newcounter{subeqncnt}
\def\thesubeqncnt{\alph{subeqncnt}}
\def\subequations{\begingroup%
\stepcounter{equation}\edef\@tempa{\theequation}%
\let\c@equation\c@subeqncnt\c@subeqncnt\z@
\edef\theequation{\@tempa\noexpand\thesubeqncnt}}

\makeatother

\slugcomment{accepted ApJ, March 14, 2008}

\shorttitle{
Coevolution of SMBHs and circumnuclear disks
}
\shortauthors{Kawakatu and Wada}

\begin{document}

\title{
Coevolution of Supermassive Black Holes 
and Circumnuclear Disks
}


\author{Nozomu Kawakatu\altaffilmark{1}}
\affil{National Astronomical Observatory of Japan, 2-21-1 Osawa, 
Mitaka, Tokyo 181-8588, Japan}

\author{Keiichi Wada}
\affil{National Astronomical Observatory of Japan, 2-21-1, Osawa, Mitaka, Tokyo 181-8588, Japan}


\altaffiltext{1}{kawakatu@th.nao.ac.jp}

\begin{abstract}
We propose a new evolutionary model of a supermassive black hole (SMBH) and
a circumnuclear disk (CND), taking into account the mass-supply from a host galaxy and the physical states of CND. In the model, two distinct accretion 
modes depending on gravitational stability of the CND play a key role on
accreting gas  to a SMBH. 
(i) If the CMD is gravitationally unstable, energy feedback from supernovae 
(SNe) supports a geometrically thick, turbulent gas disk. 
The accretion in this mode is dominated by turbulent viscosity, and it is significantly larger than that in the mode (ii), i.e.,  the CMD is supported 
by gas pressure. Once the gas supply from the host is stopped,  
the high accretion phase  ($\sim 0.01- 0.1 M_{\odot}\,{\rm yr}^{-1}$) 
changes to the low one  (mode (ii), $\sim 10^{-4} M_{\odot}\,{\rm yr}^{-1}$),
but there is a delay  with $\sim 10^{8}$ yr. 
Through this evolution, the gas-rich CND turns
into the gas poor stellar disk.  
We found that not all the gas supplied from the host galaxy 
to the central 100 pc region accrete onto the SMBH even in the high accretion phase (mode (i)), because  the part of gas is used to form stars. 
Outflow from the circumnuclear region also suppresses the growth of the SMBH.
As a result, the final SMBH 
mass ($M_{\rm BH,final}$) is not proportional to the total gas mass supplied from the host galaxy ($M_{\rm sup}$); $M_{\rm BH,final}/M_{\rm sup}$ decreases 
with $M_{\rm sup}$.
This would indicate that it is difficult to form a SMBH with $\sim 
10^{9} M_{\odot}$ observed at high-$z$ QSOs. 
The evolution of the SMBH and CND would be 
related to the evolutionary tracks of different type of AGNs. 
We found that the AGN luminosity tightly correlates with the 
luminosity of the nuclear starburst only 
in the high-accretion phase (mode (ii)). This implies that the AGN-starburst connection depends on the evolution of the AGN activity. 

\end{abstract}
\keywords{galaxies:active --- 
galaxies:nuclei --- ISM:structure --- galaxies:starburst --- 
black hole physics}

\section{Introduction}

The energy emitted by active galactic nuclei (AGNs) is commonly
ascribed to accretion onto a supermassive black hole (SMBH). 
The recent discovery of powerful QSOs at $z>$ 6 (Fan et al. 2001) implies 
that the formation of SMBHs is completed in less than 1 Gyr.  
In addition, high-resolution observations of galactic centers indicate
the presence of a SMBH whose mass correlates with the mass of 
spheroidal stellar components of galaxies, i.e., $M_{\rm BH}/M_{\rm sph}
\approx 10^{-3}$ in the nearby universe (e.g., Kormendy \& Richstone 1995; 
Marconi \& Hunt 2003). 
These indicate that the formation and evolution of a SMBH and 
a spheroidal stellar component are closely related, 
despite nine orders of magnitude difference in their size scale.

Various models have been proposed to explain the co-evolution of 
AGNs and spheroids, or the physical link between the SMBH growth and 
the starburst in spheroids (e.g., Silk \& Rees 1998; Adams et al. 2001; 
King 2003). 
However, little has been elucidated concerning the physics of 
angular momentum transfer in a spheroidal system (a bulge), 
which is inevitable for formation of SMBHs, taking into account 
the tight connection between SMBHs and spheroids. 
Umemura (2001) proposed that the $M_{\rm BH}-M_{\rm sph}$ relation 
can be explained by the mass accretion onto the galactic center from 
a galactic scale ($\sim 1\,{\rm kpc}$) via the radiation drag 
(see also Kawakatu \& Umemura 2002). 
During the hierarchical formation of a galactic bulge, the mass accretion 
due to the tidal torque driven by the major and minor merger of galaxies 
would also be an important process for SMBH formations
(e.g., Toomre \& Toomre 1972; Mihos \& Hernquist 1994; Mihos \& Hernquist 1996; Saitoh \& Wada 2004). 
However, previous studies of the co-evolution model simply postulated that 
the mass accretion onto the BH horizon, $\dot{M}_{\rm BH}$ is regulated 
by the Eddington rate, $\dot{M}_{\rm Edd}$ 
(e.g., Umemura 2001; Kawakatu, Umemura, \& Mori 2003; Granato et al. 2004), 
or equals the mass-supply rate from the host galaxy, $\dot{M}_{\rm sup}$ 
(e.g., Di Matteo et al. 2003; Kawata \& Gibson 2005; Di Matteo et al. 
2005; Springel et al. 2005; Okamoto et al. 2007).

The gas accumulated by galaxy mergers and radiation drags does not 
accrete onto a SMBH directly, since the angular momentum of 
the gaseous matter cannot be thoroughly removed. 
Thus, some residual angular momentum would terminate the 
radial infall, so the accreted gas forms a reservoir, 
i.e., a {\it circumnuclear disk}, in the central $\sim$100 pc 
around a SMBH (Fig. 1). 
In such a circumnuclear disk, the active star formation has been observed 
as the nuclear-starburst ($< 100\,{\rm pc}$) in nearby Seyfert galaxies 
(e.g., Imanishi \& Wada 2004; Davies et al. 2007; Watabe, Kawakatu \& 
Imanishi 2008). 
The nuclear starburst must affect the SMBH growth because 
radiation and/or supernova feedback due to starbursts 
can trigger the mass accretion onto a SMBH (e.g., Umemura 1997; 
Wada \& Norman 2002).
In order to understand the SMBH growth 
in terms of the evolutionary sequence of host galaxies, 
it is crucial to link mass accretion processes 
from a galactic scale with those from an accretion disk in the vicinity of 
a central BH, via the circumnuclear disk. 

To this aim, it is necessary to construct a model of 
the circumnuclear disk. Recently, Thompson et al. 2005 (hereafter T05) 
built up the radiation pressure-supported starburst disk with AGN fueling, 
considering the mass supply from host galaxies. In their model, they postulate
external torques (e.g., bars and spiral waves) in order to ensure that the gas can accrete to the galactic center before it turns into stars.
We here propose another plausible model of a nuclear starburst disk 
supported by the turbulent pressure led by supernova explosions. 
Wada \& Norman (2002; hereafter WN02) examined the structure of matter 
around AGNs associated with starburst by three-dimensional hydrodynamic 
simulations. They showed that the internal gas motion is not steady 
and the matter is inhomogeneous, but the global geometry is supported 
by internal turbulence caused by SN explosions. 
They also observed the mass accretion ($\sim 0.1M_{\odot}\,{\rm yr}^{-1}$) 
to the central a few pc region from the turbulent torus, and 
it is enhanced by the SN feedback. This means that the star formation 
and the gas accretion due to the turbulent viscosity can coexist 
in this inhomogeneous disk. 
Our model presented here relies on this picture, 
in which angular momentum is transferred by
a turbulent viscosity. 
In our model, the mass accretion process, star formation, and structure 
of the inhomogeneous circumnucler disk are naturally combined. 
We elucidate how SMBH grows from a seed BH, taking into account the mutual 
connection between the mass-supply from a host galaxy and the physical states 
of the circumnuclear disk accompanied 
by the star formation. T05, on the other hand, did not discuss a 
long-term evolution
of the circumnuclear disk and the central SMBH.

This article is arranged as follows:
In $\S 2$, we describe a new physical model for the circumnuclear 
disk supported by the turbulent pressure via SN explosions, 
considering the AGN fueling ($\S 2.1$, $\S 2.2$ and $\S 2.3$). 
In $\S 2.4$, we summarized the assumptions and 
free parameters in our model.
In $\S3$, we first show how the growth rate of a SMBH changes with time 
($\S 3.1$). Then, we examine the evolution of SMBHs and the gas and stellar 
masses in the disk ($\S 3.2$). Next, we examine the relation between 
the final mass of SMBH and the total accreted gas mass from host galaxies 
($\S 3.3$). In $\S 3.4$, we elucidate the time-evolution of AGN luminosity and 
the nuclear-starburst luminosity. By comparing with observations, 
we predict the physical states of the circumnuclear disk for 
different types of AGNs. 
Finally, we show the AGN-starburst luminosity relation, 
and then we discuss the origin of the AGN-nuclear starburst 
connection ($\S 3.5$). In $\S 4$, we discuss the effect of the AGN outflow and 
the radiation pressure caused by the young stars in disk. 
Moreover, we comment on the origin of a bias in AGN formation. 
Section 5 is devoted to our conclusions.

\section{Models}

We presuppose that the dusty gas 
is supplied around a central SMBH 
at a rate of $\dot{M}_{\rm sup}$ from a host galaxy 
whose surface density, $\Sigma_{\rm host}$, 
including the gas and stellar components, is constant in time. 
The accumulated gas forms a turbulent pressure-supported 
circumnuclear disk around a central SMBH with $M_{\rm BH}$ 
as schematically described in Fig. 1. 
In this paper, we focus on the average properties of the disk 
although we take into account the physics in the inhomogeneous 
distribution of gas in the disk as we will mention later. 
This treatment is valid because the global structure is dynamical 
stable even in the inhomogeneous distribution of gas in disk on a local scale 
(see WN02).

\subsection{Turbulent pressure-supported circumnuclear disk}
We describe the physical settings of the circumnuclear disk 
supported by the turbulent pressure via the SN explosions. 

From the radial centrifugal balance, the angular velocity $\Omega(r)$ 
is given by 
\begin{equation}
\Omega(r)^{2}=\frac{GM_{\rm BH}}{r^{3}}
+\frac{\pi G}{r}(\Sigma_{\rm disk}(r)+\Sigma_{\rm host}), 
\end{equation}
where $M_{\rm BH}$ is the mass of the BH and $\Sigma_{\rm disk}(r)$ is 
the surface density of baryonic components (the gaseous matter and stars). 

Concerning the vertical structure of circumnuclear disk, 
we assume the hydrodynamical equilibrium. 
The turbulent pressure ($P_{\rm tub}=\rho_{\rm g}v_{\rm t}^{2}$) 
associated with SN explosions is balanced to gravity, $g$ caused by 

\begin{equation}
\rho_{\rm g}(r)v_{\rm t}^{2}(r) = \rho_{\rm g}(r)gh(r), 
\end{equation}
where $v_{\rm t}(r)$ and $h(r)$ are the turbulent velocity and the scale 
height of the disk, respectively. Here, the gravity, $g$ is obtained as 
$g\equiv GM_{\rm BH}h/r^{3}+\pi G(\Sigma_{\rm disk}(r)+\Sigma_{\rm host})$. 

The geometrical thickness is determined by the balance between the 
turbulent energy dissipation and the energy input from SN explosions. 
Thus, the energy balance in unit time and volume can be obtained as 
\begin{equation}
\frac{\rho_{\rm g}(r)v_{\rm t}^{2}(r)}{t_{\rm dis}(r)}
=\frac{\rho_{\rm g}(r)v_{\rm t}^{3}(r)}{h(r)}
= \eta S_{*}(r)E_{\rm SN},
\end{equation}
where the dissipation timescale of the turbulence, $t_{\rm dis}(r)$ is assumed to be a crossing time, $h(r)/v_{\rm t}(r)$, 
$E_{\rm SN}$ is the total energy ($10^{51}\,{\rm erg}$) injected by an SN, 
$\eta$ is heating efficiency per unit mass which denotes how much energy 
from SNe is converted to kinetic energy of the matter, 
and the star formation rate per unit volume and time is 
$S_{*}(r)=C_{\rm *}\rho_{\rm g}(r)$. 
Here $C_{*}$ is the star formation efficiency and 
$\rho_{\rm g}(r)=\Sigma_{\rm g}(r)/2h(r)=f_{\rm g}\Sigma_{\rm disk}(r)/2h(r)$, 
where $\Sigma_{\rm g}$ is the surface density of gas component in the disk 
and $f_{\rm g}$ is a gas fraction. 
In the inhomogeneous distribution of gas in disk, the star formation 
can be led by gravitational collapse of higher density clumps than 
the critical density. 
Recently, Wada \& Norman (2007) showed that the star formation rate, $C_{*}$ 
is approximately proportional to the average density, $\rho_{\rm g}$
especially for a high density end, if the statistical structure of gas density 
is represented by a log-normal probability distribution.

By using eqs.(2) and (3), the turbulent velocity and the scale height 
for the region where central BH dominates the gravitational potential 
are given by 
\begin{eqnarray}
v_{\rm t}(r)&=&\left(\frac{GM_{\rm BH}}{r^{3}}\right)^{1/2}h_{\rm tub}(r), \\
h_{\rm tub}(r) &=& \left(\frac{r^{3}}{GM_{\rm BH}}\right)^{3/4}
(C_{*} \eta E_{\rm SN})^{1/2},   
\end{eqnarray} 
where $h_{\rm tub}$ denotes the scale height supported by 
the turbulent pressure due to the SNe feedback. 
We should emphasize that the validity of these solutions (eqs. (4) and (5)) 
were confirmed by the comparison with WN02. 

\vspace{5mm}
\epsfxsize=8cm 
\epsfbox{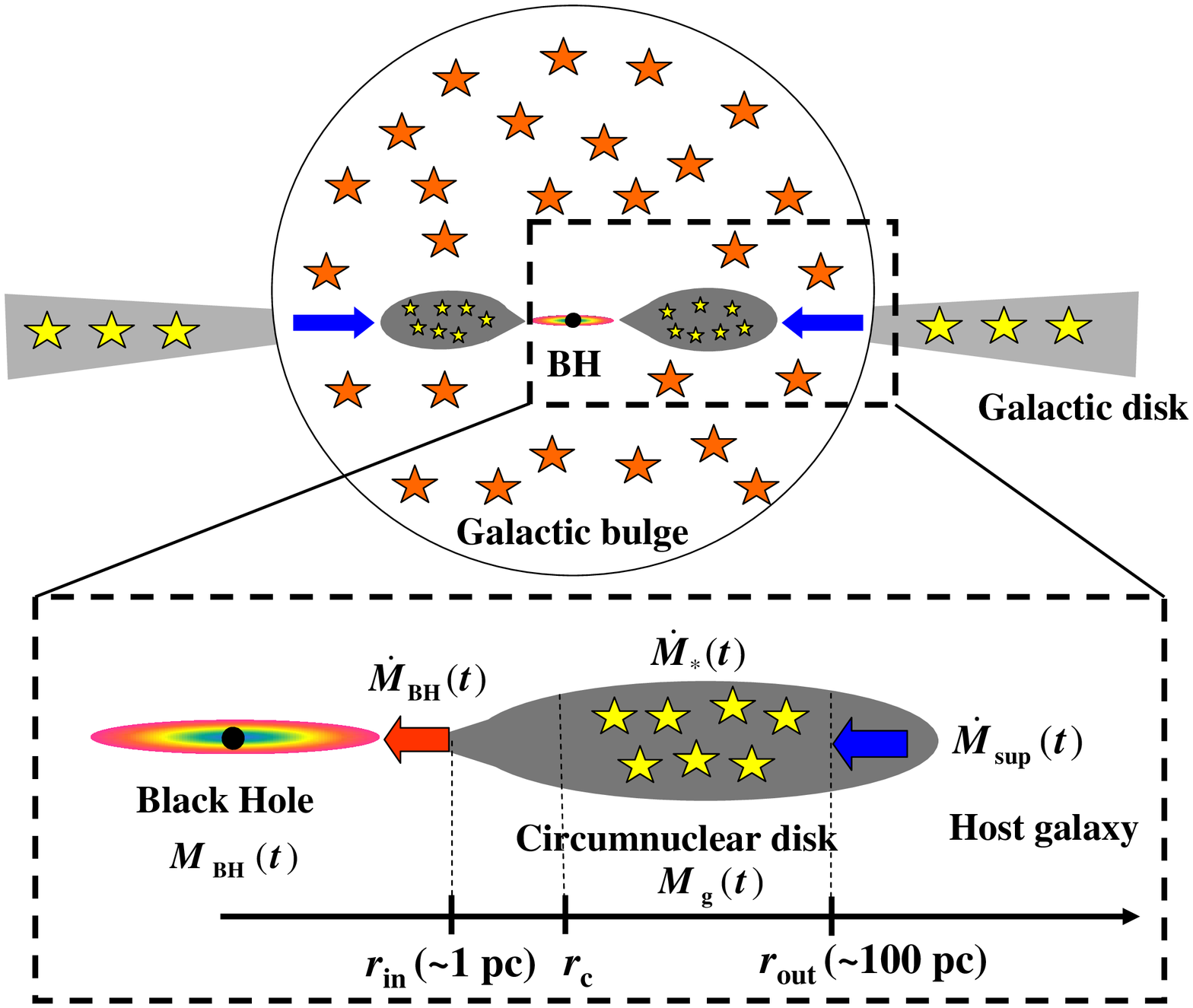}
\figcaption
{
Schematic picture for a model of SMBH growth 
linked with the change of physical state in the circumnuclear disk 
supported by the turbulent pressure via SN explosions, where 
$r_{\rm in}$ and $r_{\rm out}$ are the inner radius and the outer radius. 
Star symbols represent the region where stars can form, i.e., 
$r >r_{\rm c}$, where $r_{\rm c}$ is determined by Toomre criterion 
(see $\S 2.2$). 
Here, $M_{\rm BH}$, $M_{\rm g}$, $\dot{M}_{\rm BH}$, $\dot{M}_{*}$ 
and $\dot{M}_{\rm sup}$ are the mass of SMBHs, the gas mass, 
the stellar mass, the BH growth rate, the star formation rate, and 
the mass-supply rate from a host galaxy, respectively. 
}
\vspace{2mm}

\subsection{Two regimes of gas accretion in circumnuclear disk}
We suppose a kinetic viscosity as a source of angular momentum transfer 
in the gas disk. Then, we adopt the formula of mass accretion rate 
in a viscous accretion disk (Pringle 1981) as 

\begin{equation}
\dot{M}(r)=-\left[\frac{\frac{\partial}{\partial r}\,G(r, t)}
{\frac{d}{dr}(r^{2}\Omega(r))}\right]
=2\pi\nu\Sigma_{\rm g}(r)\left|\frac{d\,{\rm ln}\,\Omega(r)}
{d\,{\rm ln}\, r}\right|, 
\end{equation}
where the viscous torque generated by the turbulent motion, 
$G$ is defined as $G(r,t)=2\pi \nu_{\rm t}(r)\Sigma_{\rm g}(r)
r^{3}d\Omega/dr $. 
Since the viscous parameter is expressed by $\nu_{\rm t}(r)
=\alpha v_{\rm t}(r)h(r)$, the mass accretion rate in general 
increases as the energy input from SN explosions becomes large, 
in other words, the star formation rate is high 
(see eqs.(4) and (5)). 
Thus, the mass accretion rate depends on the gravitational stability 
of the disk. 
With respect to the stability, we adopt Toomre's stability criterion, 
i,e., when the surface density of gas in the disk, $\Sigma_{\rm g}$ 
is higher (lower) than the critical surface density, $\Sigma_{\rm crit}$ 
the disk is gravitationally unstable (stable). 
The critical surface density is obtained as 
\begin{equation}
\Sigma_{\rm crit}(r)=\frac{\kappa(r) c_{\rm s}}{\pi G}, 
\end{equation}
where $\kappa(r) \equiv 4\Omega(r)^{2}+2\Omega(r) rd\Omega(r)/dr$ 
is the epicyclic frequency and $c_{\rm s}$ is the sound velocity. 
The critical radius $r_{\rm c}$ is determined by the Toomre criterion, 
that is, $\Sigma_{\rm g}(r_{\rm c})=\Sigma_{\rm crit}(r_{\rm c})$. 

In this picture, it is natural that there are two modes of gas accretion rate 
as follows: 
\underline{mode (i):} If the circumnuclear disk is fully gravitationally unstable, in other words, the critical radius for the gravitational instability, $r_{\rm c}$ is smaller than the inner radius of the disk, $r_{\rm in}$, then the disk is geometrically thick due to stellar energy feedback, and as a result we have a large accretion rate. The value of $\alpha$ produced by randomly distributed SN explosions is not obvious. However, thanks to numerical simulations 
demonstrated by WN02, the average transfer of angular momentum 
can be described as $\nu_{\rm t}(r)\approx v_{\rm t}(r)h(r)$. 
Thus, we suppose $\alpha=O(1)$. 
\underline{mode (ii):} If the critical radius is located inside the disk 
($r_{\rm c} > r_{\rm in}$), the scale height of the inner region would 
be much smaller than mode (i), because the scale height is determined by 
the thermal pressure, 
$P_{\rm g}(r)=\rho_{\rm g}(r)gh_{\rm th}(r)$ 
where $P_{\rm g}(r)=\rho_{\rm g}(r)c_{\rm s}^{2}$. 
In the mode (ii),  the magneto-rotational instability could be a source of 
turbulence, but the turbulent velocity is comparable or even smaller than 
the sound speed, i.e., $\alpha \approx 0.01-0.5$ and $v_{\rm t}=c_{\rm s}$ 
(e.g., Balbus \& Hawley 1991; Machida et al. 2000; Machida \& Matumoto 2003).
As a result, the accretion is less efficient than the mode (i).

Since the critical radius and the inner edge of the disk are functions of 
the BH mass, and surface density of the gas, and the gas mass depends on 
the star formation rate in the gas disk, the evolution of 
the whole system (a central BH plus a circumnuclear disk) should be time-dependent. 

\subsection{SMBH growth and states of the circumnuclear disk}
Based on the picture described in $\S 2.1$ and $\S2.2$. 
we here examine the evolution of central BHs and circumnuclear disk, 
considering the mass supply from host galaxies (see Fig. 1). 
For this purpose, we evaluate the time-evolution of SMBH growth and 
the star formation rate and gas mass in the disk by focusing on 
the time dependence of characteristic radius in the disk, 
instead of solving the evolution of radial structure of disk. 
Then, the surface density of disk is assumed by a power law 
with the cylindrical radius $r$ as 
$\Sigma_{\rm disk}(r)=\Sigma_{\rm disk,0}(r/r_{\rm out})^{-\gamma}$, 
where $r_{\rm out}$ is the outer boundary of the disk, 
which will be defined below. 
Hereafter the physical quantities with suffix 0 are the value 
at $r_{\rm out}$. 

In this model the supplied gas from the host galaxy 
is eventually consumed to form the SMBH or stars 
(see $\S 4.1$ for the effect of AGN outflow).
Then, the BH mass and gas mass in the disk are simply given 
by the mass conservation 
as follows: 
The time-evolution of the gas mass in the disk, 
$M_{\rm g}\equiv\int^{r_{\rm out}}_{r_{\rm in}} 2\pi r^{\prime}
\Sigma_{\rm g}(r^{\prime})dr^{\prime}$, is given by
\begin{equation}
M_{\rm g}(t)=\int^{t}_{0}[\dot{M}_{\rm sup}(t^{\prime})
-\dot{M}_{*}(t^{\prime})
-\dot{M}_{\rm BH}(t^{\prime})]dt^{\prime}, 
\end{equation}
where $\dot{M}_{\rm sup}(t)$, $\dot{M}_{*}(t)$ 
and $\dot{M}_{\rm BH}(t)$ are the mass-supply rate from hosts, 
the star formation rate, 
and the growth rate of SMBH, respectively. 
On the other hands, the time-evolution of SMBH mass $M_{\rm BH}(t)$ 
is obtained as 
\begin{equation}
M_{\rm BH}(t)=M_{\rm BH,seed}+\int^{t}_{0}\dot{M}_{\rm BH}
(t^{\prime})dt^{\prime}, 
\end{equation}
where we assume the mass of seed BHs, $M_{\rm BH,seed}=10^{3}M_{\odot}$, 
as end-products of the first generation stars (e.g., Heger et al 2003). 

As for $\dot{M}_{\rm sup}$, we can assume any function for the mass supply rate, but here we simply take a step function as the first attempt as 
\begin{equation}
\dot{M}_{\rm sup}(t)=\left \{
 \begin{array}{l}
 const.,\,\,\ {\rm for}\,\,\, t < t_{\rm sup}, \\
 0,\,\,\ {\rm for}\,\,\, t \ge t_{\rm sup}, 
 \end{array}
 \right .
\end{equation}
where $t_{\rm sup}$ is a period of the mass-supply from hosts. 

The star-formation rate in the disk can be defined as 
$\dot{M}_{*}\equiv 
2\int^{r_{\rm out}}_{r_{\rm min}} S_{*}(r^{\prime})h(r^{\prime})
2\pi r^{\prime} dr^{\prime}$, 
where $r_{\rm min}$ is the minimum radius at which stars can form. 
Here $r_{\rm min}=r_{\rm in}$ for $r_{\rm c} < r_{\rm in}$ and 
$r_{\rm min}=r_{\rm c}$ for $r_{\rm c} > r_{\rm in}$. 
Thus, $\dot{M}_{*}(t)$ can be expressed in terms of 
$M_{\rm g}(t)$ as follows: 
\begin{equation}
\dot{M}_{*}(t) = AC_{*} M_{\rm g} (t), 
\end{equation}
where $A=1$ for $r_{\rm c} < r_{\rm in}$, while 
$A=1-(r_{\rm c}/r_{\rm out})^{3-\gamma}$
for $r_{\rm c} > r_{\rm in}$. 

We define the growth rate of SMBH, $\dot{M}_{\rm BH}$ 
as $\dot{M}_{\rm BH}\equiv \dot{M}(r_{\rm in})$, 
where $\dot{M}(r_{\rm in})$ is a mass accretion rate in the inner region 
of the circumnuclear disk.
By using eq. (6), the BH growth rate is calculated for 
$0 \leq \gamma < 2$ as follows:

\begin{equation} 
\dot{M}_{\rm BH}=\left \{
 {\small
 \begin{array}{l}
 f_{\rm b}\eta E_{\rm SN}(GM_{\rm BH}/r_{\rm out})^{-1}
 \dot{M}_{*}
 (r_{\rm in}/r_{\rm out})^{3-\gamma},\,\,\, 
 {\rm for}\,\,\, {\rm mode\,(i)}, \\ 
 \\
  3f_{\rm geo}(\alpha c_{\rm s}^{3}/G)
  (M_{\rm g}/M_{\rm BH})
 (r_{\rm in}/r_{\rm out})^{2-\gamma}, \,\,\,
 {\rm for}\,\,\,, {\rm mode\,(ii)}
 \end{array}
 }
 \right .
\end{equation}

where $f_{\rm b}\equiv 3(14-4\gamma)(8-4\gamma)/16$ and 
$f_{\rm geo}\equiv r_{\rm in}/h(r_{\rm in})$ 
\footnote{
Let us consider $r_{\rm in}$, $r_{\rm out}$, $M_{\rm g}$, 
$M_{*}$ and $M_{\rm BH}$ dependences on $\dot{M}_{\rm BH}$ 
(see eq.(12)). 
As smaller $r_{\rm out}$ or larger $r_{\rm in}$, $\Sigma_{\rm g}(r_{\rm in})$ 
increases because of $\Sigma_{\rm g}(r)\propto r^{-\gamma}$ 
($0\le \gamma < 2$). 
The larger $r_{\rm in}$ (or smaller $r_{\rm out}$) leads to 
larger $\dot{M}_{\rm BH}$, 
because $\dot{M}_{\rm BH}\propto \Sigma_{\rm g}(r_{\rm in})$ from eq. (6).
The larger BH mass leads to a smaller scale height from 
eq. (5). Thus, $\dot{M}_{\rm BH}$ becomes smaller 
since $\dot{M}_{\rm BH}\propto h$.  
As $\dot{M}_{*}\,(\propto M_{\rm g})$ and $M_{\rm g}$ 
are larger, $\Sigma_{\rm g}(r_{\rm in})$ is larger, and thus 
$\dot{M}_{\rm BH}$ increases.}
.

\subsection{Assumptions and free parameters}
In this section, we summarize the assumptions and free parameters 
as follows;

\hspace*{-0.3cm}\underline{Surface density of hosts: $\Sigma_{\rm host}$}\\
We fix the surface density of host galaxies 
with $\Sigma_{\rm host}=10^{4}M_{\odot}\,{\rm pc}^{-2}$. 
This is the upper value of nearby starburst galaxies (e.g., 
Kennicutt et al. 1998), but may be a typical value of high-$z$ 
starburst galaxies (e.g., Tacconi et al. 2006). 

\hspace*{-0.3cm}\underline{Surface density of disk: $\Sigma_{\rm disk}(r)$} \\
As for the power law surface density profile of the disk, we should note 
that recent numerical simulations of AGN disks showed 
that the single power law profile is maintained (e.g., Levine et al. 2007). 
Such a power low profile would be supported because 
the turbulence generated by the self-gravity and/or the SN feedback 
redistributes the angular momentum in the disk on a dynamical time scale. 
In this paper, we change the power law index $\gamma$ from 0 to 2. 

\hspace*{-0.3cm}\underline{Gas temperature of disk: $T_{\rm g}$} \\
As for the gas temperature, we assume the isothermal gas with 
$T_{\rm g}=100\, {\rm K}$, that is, $c_{\rm s}\simeq1\,{\rm km}{\rm s}^{-1}$.
Recent numerical simulations showed that the circumnuclear disk around a 
SMBH would be dominated by the cold gas ($T_{\rm g} \approx 50-100$ K) since the dust cooling is effective (e.g., Wada \& Tomisaka 2005; see also Levine et al. 2007). In addition, the gas temperature is almost constant for the 
range of gas density in circumnuclear disks (Fig. 5 in Wada \& Tomisaka 2005). 
These indicate that an isothermal gas with $T_{\rm g}=100\, K$ 
is reasonable assumption for discussing mass evolution of the 
circumnuclear disk. 

\hspace*{-0.3cm}\underline{Outer radius: $r_{\rm out}$} \\
The size $r_{\rm out}$ is defined as the outer boundary 
inside which the potential of the BH plus circumnuclear disk system dominates 
that of the host galaxy. 
From Toomre condition, 
the outer part of the disk ($r > r_{\rm out}$) is always 
gravitationally stable, and then no stars can form in the disk. 
Since our main aim is to reveal the physical connection 
between the AGN activity (the SMBH growth) and the starburst in the 
circumnuclear region, 
we concentrate on inner parts of the circumnuclear disk 
($ r < r_{\rm out}$). Then, when $M_{\rm disk} > M_{\rm BH}$ 
the outer radius $r_{\rm out}$ is given by 
$GM_{\rm disk}/r_{\rm out}=GM_{\rm BH}/r_{\rm out}
+\pi G\Sigma_{\rm host}r_{\rm out}$, 
where $M_{\rm disk}\equiv \int_{r_{\rm in}}^{r_{\rm out}} 
2\pi r^{\prime}\Sigma_{\rm disk}(r^{\prime})dr^{\prime}$, 
and $r_{\rm in}$ is the inner radius of the disk. 
From eq. (2), $r_{\rm out}$ is 
$\approx (M_{\rm disk}/\pi \Sigma_{\rm host})^{1/2}
\sim 60\, {\rm pc}\, M_{\rm disk, 8}^{1/2}\Sigma_{\rm host, 4}^{-1/2}$, 
where $M_{\rm disk, 8}$ and $\Sigma_{\rm host,4}$ 
are the total baryonic mass of the circumnuclear disk normalized 
by $10^{8}M_{\odot}$ 
and the surface density of host galaxy normalized by 
$10^{4}M_{\odot}\,{\rm pc}^{-2}$, respectively. 
On the other hand, if $M_{\rm disk} < M_{\rm BH}$, 
$r_{\rm out}$ is obtained from $GM_{\rm BH}/r_{\rm out}
=\pi G\Sigma_{\rm host}r_{\rm out}$.
From this, we quantitatively evaluate $r_{\rm out}$ as 
$r_{\rm out}=(M_{\rm BH}/\pi \Sigma_{\rm host})^{1/2}
\sim 60\, {\rm pc}\, M_{\rm BH, 8}^{1/2}\Sigma_{\rm host, 4}^{-1/2}$, 
where $M_{\rm BH,8}$ is the mass of BH normalized by $10^{8}M_{\odot}$.

\hspace*{-0.3cm}\underline{Inner radius: $r_{\rm in}$} \\
We suppose that the inner radius of the disk, $r_{\rm in}$, 
is determined by the dust-sublimation radius, 
which is strongly supported by observational results 
(e.g., Suganuma et al. 2006).
Thus, the inner radius $r_{\rm in}$ is expressed as 
$r_{\rm in}=3\,{\rm pc}\, L_{\rm AGN, 46}^{1/2}T_{1500}^{-2.8}$, 
where $L_{\rm AGN, 46}$ is the AGN luminosity in units of $10^{46}{\rm erg}\,
{\rm s}^{-1}$ and $T_{1500}$ is the silicate grains sublimation temperature 
in units of 1500 K (Laor \& Draine 1993). 
Although $r_{\rm in}$ depends on the size and composition of dust 
grains (e.g., Barvainis 1992; Sitko et al. 1993; Laor \& Draine 1993; 
Kishimoto et al. 2007), our conclusions does not change so much. 
Provided that $L_{\rm AGN}$ is equal to the Eddington luminosity, 
$L_{\rm Edd}=4\pi cGM_{\rm BH}m_{\rm p}/\sigma_{\rm T}$, 
the inner radius can be rewritten as 
$r_{\rm in}=3\,{\rm pc}\,M_{\rm BH, 8}^{1/2}T_{1500}^{-2.8}$, 
where $m_{\rm p}$ is the proton mass and 
$\sigma_{\rm T}$ is the Thomson cross-section. 
In this paper, we adopt $r_{\rm in}=3\,{\rm pc}\,M_{\rm BH, 8}^{1/2}$. 

In our model, we have three free parameters as follows: 
(i) the power law index of surface density, $\gamma$, 
(ii) the heating efficiency, $\eta$ and 
(iii) the star formation efficiency, $C_{*}$. 
In the following sections, we use the fiducial value as 
$\gamma=1$, $\eta=10^{-3}M_{\odot}^{-1}$ which can be derived 
from comparing our analytical solution of scale height with 
numerical simulations done by WN02 and 
$C_{*}=3\times10^{-8}\,{\rm yr}^{-1}$ 
which is upper value of nearby starburst galaxies 
(Kennicutt 1998; Fig.13 in Wada \& Noman 2007). 
However, since $\gamma$, $\eta$ and $C_{*}$ are free parameters in this work 
we will discuss the effect of changing these values in $\S 3.2$.
For these fiducial free parameters, 
$\dot{M}_{\rm BH}$ (eq. (12)) can be evaluated quantitatively 
as follows:
\begin{equation}
\frac{\dot{M}_{\rm BH}(t)}{M_{\odot}\,{\rm yr}^{-1}}= \left \{
 \begin{array}{l}
 0.3 \,\eta_{-3}C_{*, -8}\Sigma_{\rm host,4}^{1/2}M_{\rm g, 8}(t)
 M_{\rm disk, 8}(t)^{-1/2},\,\,\, 
\\{\rm for\,\,\, mode\,(i)}\\
 3\times 10^{-3}\, \alpha_{0.5}\, 
 c_{\rm s,1}^{3} \, \Sigma_{\rm host,4}^{1/4}\, M_{\rm g, 8}(t)
\,M_{\rm disk,8}(t)^{-1/2}\\ \times M_{\rm BH,8}(t)^{-1/4},\,\,\, 
{\rm for\,\,\, mode\,(ii)}, 
 \end{array}
 \right .
\end{equation}
where 
$\eta_{-3}$ is the heating efficiency normalized by 
$10^{-3}M_{\odot}^{-1}$, 
$C_{*,-8}$ is the star formation efficiency normalized by 
$10^{-8}\,{\rm yr}^{-1}$, $M_{\rm g, 8}$ is gas mass of 
massive disk normalized by $10^{8}M_{\odot}$, 
$\alpha_{0.5}$ is the $\alpha$-parameter normalized by 0.5, 
and $c_{\rm s,1}$ is the sound velocity normalized 
by $1\,{\rm km}\,{\rm s}^{-1}$. 

\section{Results}
\subsection{The growth rate  of SMBHs and star formation rate}
Based on the physical model connecting between the growth of SMBHs 
and the physical states of circumnuclear disks described in 
$\S 2$, we examine how the growth rate of SMBHs is related to 
the star formation rate in the circumnuclear disk.

Figure 2 shows the time evolution of $\dot{M}_{\rm BH}(t)$, 
$\dot{M}_{*}(t)$, and $\dot{M}_{\rm Edd}(t)\equiv L_{\rm Edd}(t)/c^{2}$ 
for $\dot{M}_{\rm sup}=1\, M_{\odot}\,{\rm yr}^{-1}$ and 
$t_{\rm sup}=10^{8}\,{\rm yr}$. 
The critical time-scale, $t_{\rm crit}$ 
is the time when $r_{\rm c}=r_{\rm in}$. 
Thus, the stars can form in the whole region of the disk 
before $t = t_{\rm crit}$. 
We define the former phase ($t < t_{\rm crit}$) 
as the high-accretion phase and 
the latter phase ($t > t_{\rm crit}$) as the low-accretion phase. 
As seen in Fig. 2, we find that not all the gas supplied from a 
host galaxy reaches to the SMBH, 
i.e., $\dot{M}_{\rm BH}=0.1-0.3M_{\odot}\,{\rm yr}^{-1}$ 
for $\dot{M}_{\rm sup}=1\, M_{\odot}\,{\rm yr}^{-1}$ 
even in the high accretion phase ($r_{\rm c} < r_{\rm in}$). 
This high accretion phase drastically changes to the low 
accretion phase with $\dot{M}_{\rm BH}\sim 10^{-4}
M_{\odot}\,{\rm yr}^{-1}$. 
But, there is a delay with $\sim 10^{8}\,{\rm yr}$ 
after the mass supply is stopped. 
The disk becomes gravitationally stable because the surface density 
of gas in the disk decreases due to the starformation. 
Thus, this timescale is basically determined by the star formation 
timescale, $t_{*}=C_{*}^{-1}\sim 3\times 10^{7}\,{\rm yr}$.
The drastic change of $\dot{M}_{\rm BH}$ depends on whether stars 
are formed in the inner region of circumnuclear disk, because 
the ratio of $\dot{M}_{\rm BH}$ between high and low accretion phase 
is given by 
\begin{eqnarray}
\frac{\dot{M}_{\rm BH}({\rm for}\,\,\, r_{\rm c}<r_{\rm in})}
{\dot{M}_{\rm BH}({\rm for}\,\,\, r_{\rm c}>r_{\rm in})}
&=&\frac{v_{\rm t}h_{\rm tub}}{\alpha c_{\rm s}h_{\rm th}}
\sim 10^{3}, \nonumber 
\end{eqnarray}
where $v_{\rm t}/(\alpha c_{\rm s})\approx 10^{2}M_{\rm BH,8}^{1/8}$ 
and $h_{\rm tub}/h_{\rm th}\approx 10M_{\rm BH,8}^{1/8}$.

\vspace{5mm}
\epsfxsize=8cm 
\epsfbox{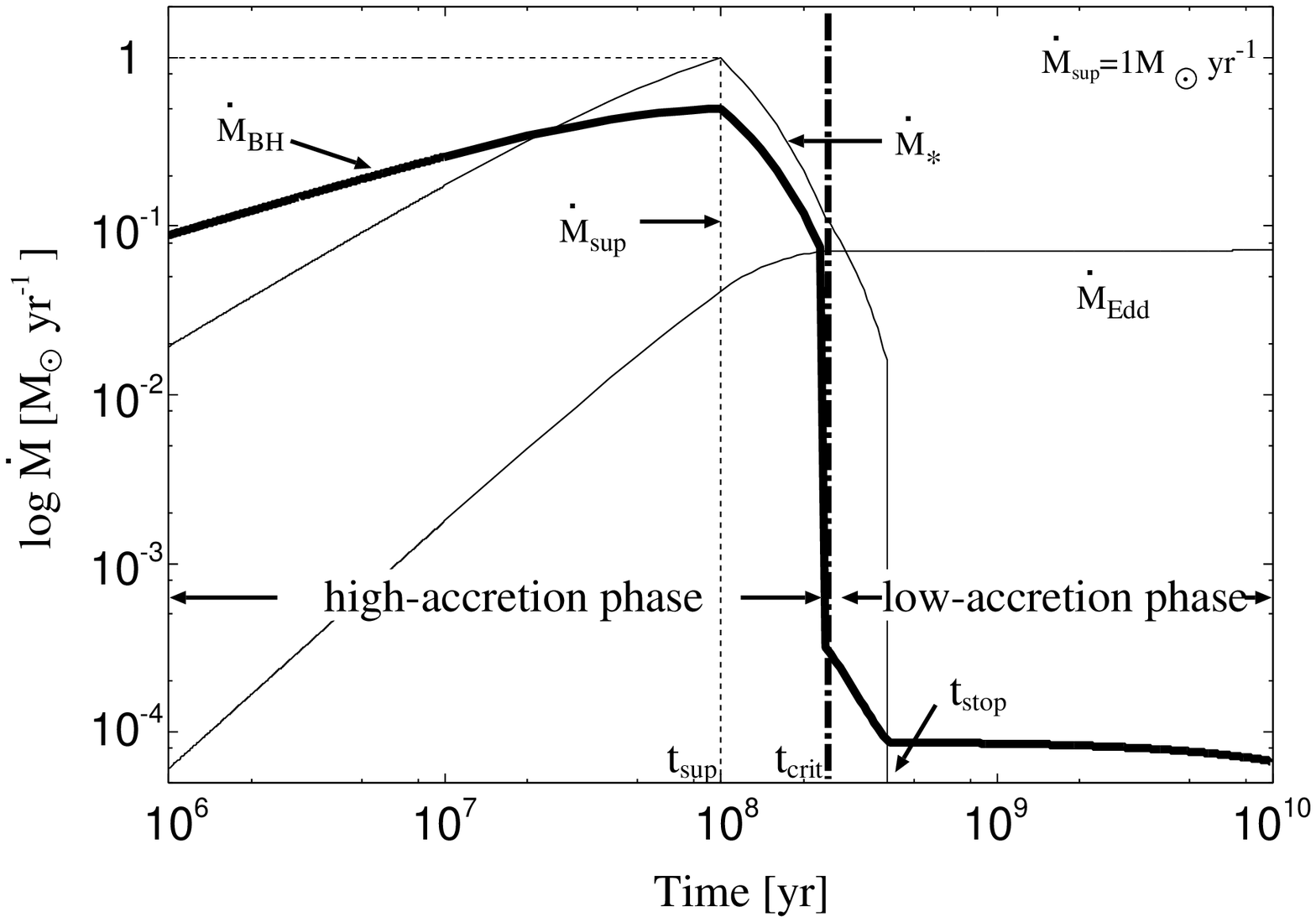}
\figcaption
{
Time evolution of $\dot{M}_{\rm BH}(t)$, $\dot{M}_{*}(t)$, 
$\dot{M}_{\rm sup}(t)$ and $\dot{M}_{\rm Edd}(t)$ 
for $\dot{M}_{\rm sup}=1\,M_{\odot}\,{\rm yr}^{-1}$ and 
$t_{\rm sup}=10^{8}\,{\rm yr}$. 
$t_{\rm sup}$ is a period of the mass-supply from hosts. 
$t_{\rm crit}$ is the time when $r_{\rm c}=r_{\rm in}$. 
$t_{\rm stop}$ is the time when the massive disk is gravitationally 
stable, that is, $\Sigma_{\rm g} < \Sigma_{\rm crit}$ for the entire 
massive disk. 
We call the phase ($t < t_{\rm crit}$) the high-accretion phase, 
while we call the phase ($t > t_{\rm crit}$) the low-accretion phase. 
The star formation efficiency $C_{*}$ is $3\times 10^{-8}\,{\rm yr}$.
}
\vspace{2mm}

It also has been found that SMBH growth dominates 
the star formation in the disk (i.e., $\dot{M}_{\rm BH} > \dot{M}_{\rm *}$) 
in the early epoch, i.e., $t < t_{*}=C_{*}^{-1}\sim 3\times 10^{7}\,{\rm yr}$. 
In this phase, a super-Eddington mass accretion rate ($\dot{M}_{\rm BH} \gg 
\dot{M}_{\rm Edd}$) keeps for $\sim 10^{8}\,{\rm yr}$. 
Recent two-dimensional radiation hydrodynamic simulations have shown 
that such a super-Eddington mass accretion flow is possible 
(e.g., Ohsuga et al. 2005). 
After this phase, $\dot{M}_{*}$ is larger than $\dot{M}_{\rm BH}$ 
because of $\dot{M}_{*}\propto t$ and $\dot{M}_{\rm BH}\propto t^{1/2}$
\footnote{During $t < t_{\rm sup}$, $\dot{M}_{\rm BH}\propto M_{\rm g}^{1/2}$ 
because $M_{\rm disk}$ is nearly equal to $M_{\rm g}$ from eq. (12). 
On the other hand, $\dot{M}_{*}\propto M_{\rm g}$ from eq. (11),  
considering $M_{\rm g}\propto t$ until $t=t_{\rm sup}$ 
(see Fig. 3).}. 
Then, the SMBH growth rate is much smaller than the Eddington mass accretion 
rate because of the gas pressure supported geometrically thin disk. 
After $t=t_{\rm stop}$ the star formation terminates because the entire 
disk is gravitationally stable, where $t_{\rm stop}$ is the time when 
$\Sigma_{\rm g}(r)$ is smaller than $\Sigma_{\rm crit}(r)$ 
throughout the entire circumnuclear disk.

\subsection{Evolution of gas, SMBH and stellar masses}
In this section, we elucidate the physical difference between two phases for 
SMBH growth (high- and low- accretion phases). 
We plot the mass of SMBH, $M_{\rm BH}(t)$, gas mass in the 
disk, $M_{\rm g}(t)$, and stellar mass in the 
disk, $M_{*}(t)\equiv \int^{t}_{0}\dot{M}_{*}(t^{\prime})dt
^{\prime}$ in Fig. 3. 
The total supplied gas mass from the hosts, $M_{\rm sup}\equiv 
\dot{M}_{\rm sup}t_{\rm sup}$, is $10^{8}\,M_{\odot}$. 
In the high accretion phase, plenty of gas accumulate around a SMBH 
since $\dot{M}_{\rm sup}(t) > \dot{M}_{\rm BH}(t)$ (see Fig. 2). 
Thus, the mass ratio of gas in circumnuclear disk and SMBH, 
$m_{\rm disk}\equiv M_{\rm g}/M_{\rm BH}$ is about ten 
and the gas fraction to the total baryonic mass, $f_{\rm g}\equiv 
M_{\rm g}/(M_{\rm g}+M_{\rm *})$, is close to $\approx 1$. 
Recalling that the accretion is a super-Eddington in high accretion phase, 
large $m_{\rm disk} > 1$ and $f_{\rm g}\sim 1$ would be 
the conditions which super-Eddington mass accretion can keep. 
Since the scale height in the inner region is inflated by the energy input 
led by the SN explosions, 
the circumnuclear disk would be geometrically thick, 
which may correspond to the obscuring torus proposed by the AGN 
unified model (e.g., Antoucci 1993; Urry \& Padovani 1995; WN02). 
On the other hand, in the low-accretion phase ($t > t_{\rm crit}$), 
stellar mass dominates the BH mass and gas mass, 
i.e., $f_{\rm g}\approx 10^{-2}$ and $m_{\rm disk}\approx 0.03$. 
In this phase, the disk is the gravitationally stable, so that 
the inner region of disk should be geometrically thin. 

Moreover, we find that the saturation of SMBH growth 
appears $\sim 10^{8}\,{\rm yr}$ later after the mass supply is stopped. 
One should note that the final BH mass is about 1/3 of the total gas mass 
supplied from hosts. 
The timescale of the SMBH growth, $t_{\rm growth}$ is determined 
by the viscous timescale, $t_{\rm vis}$ in the disk as follows: 
\begin{eqnarray}
\frac{t_{\rm vis}}{10^{8}\,{\rm yr}}
\approx \frac{r_{\rm in}^{2}}{v_{\rm t}h}=
\frac{GM_{\rm BH, final}}{\eta E_{\rm SN}r_{\rm in}}C_{*}^{-1}
\sim 5\left(\frac{\eta}{10^{-3}M_{\odot}^{-1}}\right)^{-1}\\ \nonumber
\times \left(\frac{M_{\rm BH,final}}{10^{8}M_{\odot}}\right)^{1/2}
\left(\frac{t_{*}}{10^{8}\,{\rm yr}}\right),
\end{eqnarray}
where $M_{\rm BH, final}$ is the final SMBH mass. 
At given $\eta=10^{-3}M_{\odot}^{-1}$ 
and $M_{\rm BH,final}=3\times 10^{7}M_{\odot}$ (see Fig.3), 
the timescale of SMBH growth is comparable to that of star formation 
in the gas disk. 
From eq. (14), it is found that the rapid BH growth 
(i.e., $t_{\rm growth} < 10^{9}\,{\rm yr}$) 
can be achieved if the star formation rate in the disk is quite high, 
e.g., $t_{*} \leq 10^{8}\,{\rm yr}$. 
Hence, it is essential to reveal what determines the star formation timescale 
(i.e.,  the star formation efficiency, $C_{*}$) in the circumnuclear disk. 
This will be left as our future works. 
For the dependence of $\eta$ and $M_{\rm BH, final}$, we can understand 
as follows. 
From eq. (5), the larger BH mass leads to a smaller scale height, 
so that $t_{\rm growth}$ becomes larger as $M_{\rm BH,final}$ increases 
because of $\dot{M}_{\rm BH}\propto h$. 
Since $\dot{M}_{\rm BH}\propto \eta$ (see eq. (12)), $t_{\rm growth}$ 
is longer as $\eta$ becomes larger.

Finally, we discuss the effect of three free parameters ($\gamma$, $C_{*}$
and $\eta$). 
As for $\gamma$, we examine $\dot{M}_{\rm BH}$ for $\gamma=0.5$ 
and $\gamma =1.5$ by using the analytical solutions (see eq. (12)), 
and find that our results do not depend on the difference of 
$\gamma$ significantly, that is, 
the difference in $\dot{M}_{\rm BH}$ between 
$\gamma=1$  and $\gamma=0.5$ (or 1.5) is less than a 
factor of 2 although $\dot{M}_{\rm BH}$ is larger as 
larger $\gamma$. 
For $C_{*}$, both the star formation rate and 
and BH growth rate in the high accretion phase 
are ten times smaller from eqs. (11) and (12), 
if we adopt ten time smaller $C_{*}$, i.e., 
$C_{*}=3\times 10^{-9}\,{\rm yr}^{-1}$. 
Thus, after $t_{\rm sup}$ the gas mass in the disk decreases slower 
(see eq. (8)), and then $t_{\rm crit}$ and $t_{\rm stop}$ become roughly 
ten times longer. However, the final BH mass ($M_{\rm BH,final}\approx 
\dot{M}_{\rm BH}t_{\rm growth}$) do not change significantly 
because of $\dot{M}_{\rm BH}\propto C_{*}$ and $t_{\rm growth}\propto 
C_{*}^{-1}$.  
This is confirmed by Fig. 4, which is the same as Fig. 2 and Fig. 3 but 
the star formation efficiency is ten times smaller 
(i.e., $C_{*}=3\times 10^{-9}\,{\rm yr}^{-1}$). 
Concerning $\eta$, if $\eta$ is ten times smaller than the value we adopted, 
i.e., $10^{-4}M_{\odot}^{-1}$, the final BH mass is ten times smaller 
because of $\dot{M}_{\rm BH}\propto \eta$ in the high accretion phase 
(see eq. (12)).

\vspace{5mm}
\epsfxsize=8cm 
\epsfbox{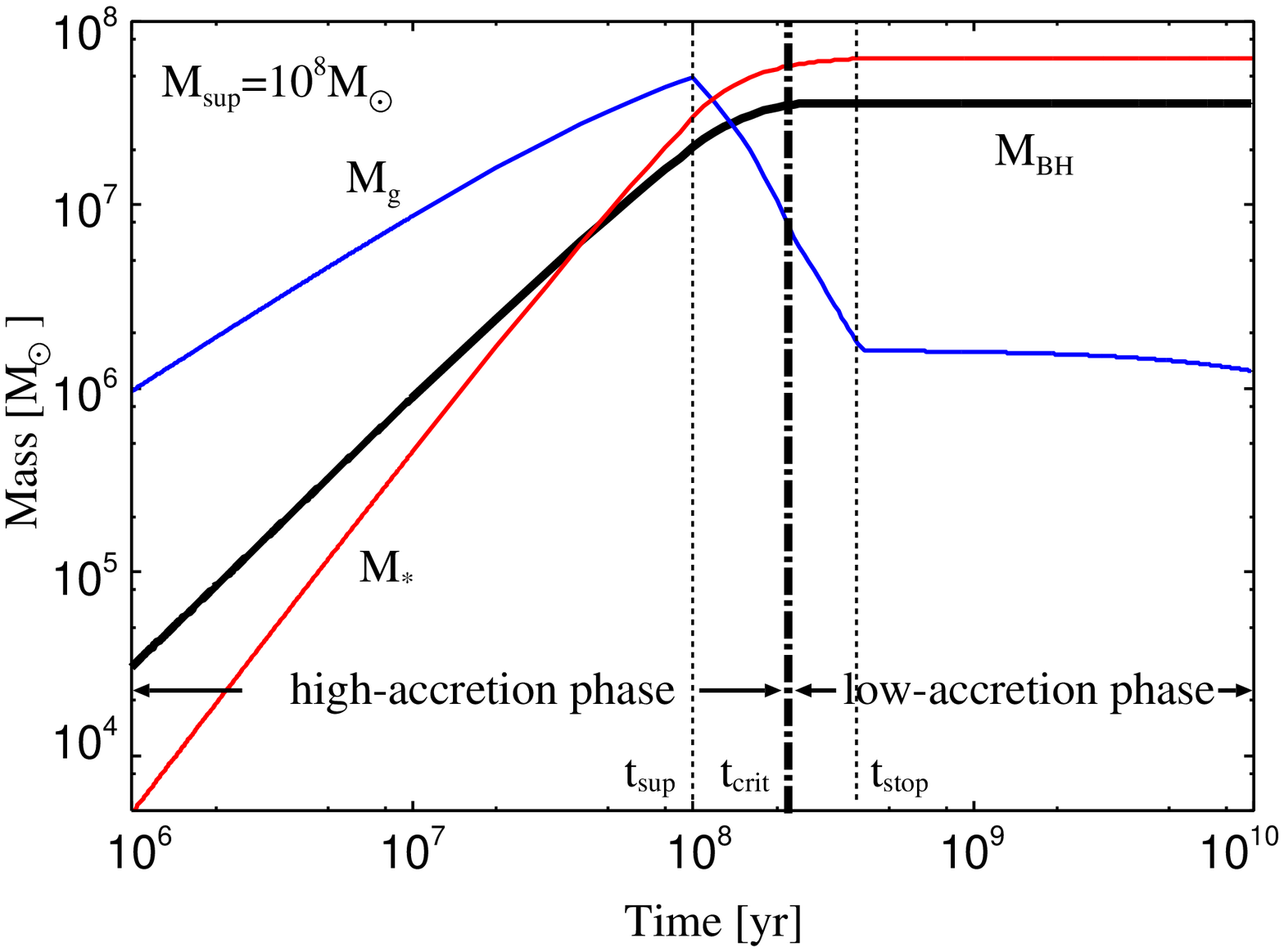}
\figcaption
{
Time evolution of the mass of BH (black solid line), $M_{\rm BH}(t)$, 
the gas mass in the disk (blue solid line), $M_{\rm g}(t)$, 
and the stellar mass in the disk (red solid line), $M_{*}(t)$. 
The total supplied mass from host galaxies $M_{\rm sup}$ 
is $10^{8}\,M_{\odot}$. The star formation efficiency $C_{*}$ is 
$3\times 10^{-8}\,{\rm yr}$. Note that $m_{\rm disk}\equiv 
M_{\rm g}/M_{\rm BH}$ and $f_{\rm g}\equiv M_{\rm g}/(M_{\rm g}+M_{*})$.
}
\vspace{2mm}

\vspace{5mm}
\epsfxsize=8cm 
\epsfbox{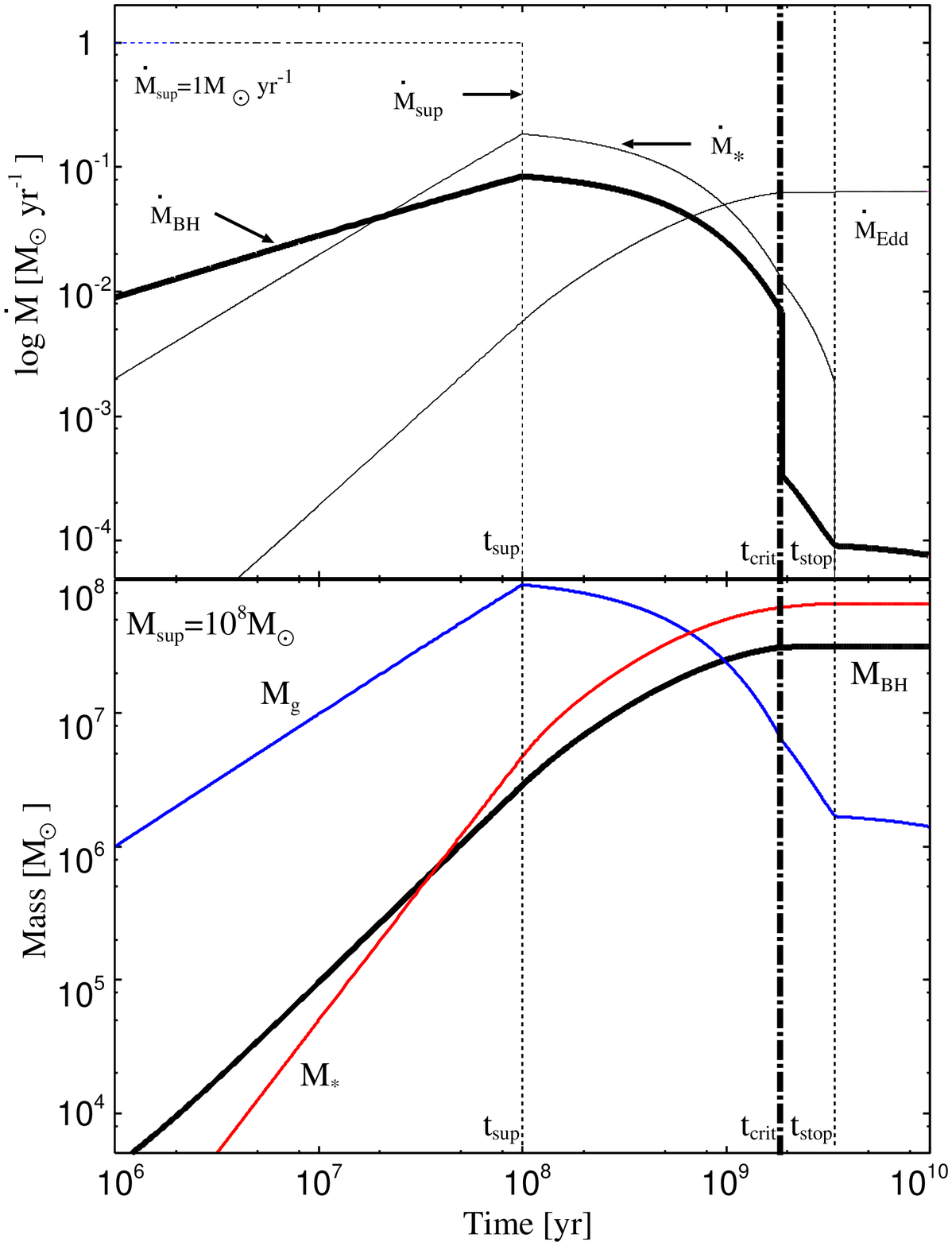}
\figcaption
{
Same as Fig. 2 and Fig. 3, but the star formation efficiency $C_{*}$ 
is ten times smaller (i.e., $C_{*}=3\times 10^{-9}\,{\rm yr}^{-1}$). 
}
\vspace{2mm}

\subsection{$M_{\rm BH, final}$-to-$M_{\rm sup}$ relation}
As mentioned in $\S3.2$, the final mass of SMBH, 
$M_{\rm BH, final}$, is not equal to $M_{\rm sup}$, 
i.e., $M_{\rm BH, final}\approx 0.3M_{\rm sup}$. 
In this section, we examine the relation between $M_{\rm BH, final}$ and 
$M_{\rm sup}$. 
Figure 5 shows the final BH mass as a function of  total accreted mass 
from host galaxies with $M_{\rm sup}=10^{6}M_{\odot}-10^{10}M_{\odot}$. 
We here assume $t_{\rm sup}=10^{8}\,{\rm yr}$. 
Since the solid line is our prediction and the dot-dashed line denotes 
$M_{\rm BH,final}=M_{\rm sup}$, it is found that 
$M_{\rm BH, final}/M_{\rm sup}$ gets smaller 
as $M_{\rm sup}$ becomes larger, i.e., 
$M_{\rm BH, final}/M_{\rm sup}=0.8, \,0.6,\, 0.3,\, 0.1$ and $0.05$ 
for $M_{\rm sup}=10^{6}, 10^{7}, 10^{8}, 10^{9}$, and 
$10^{10}M_{\odot}$, respectively. 
In other words, the final mass of SMBHs, $M_{\rm BH, final}$ 
is not proportional to $M_{\rm sup}$ and growth of a SMBH is 
more inefficient for larger $M_{\rm sup}$.  
As mentioned in $\S 3.1$ and $\S 3.2$, the star formation in the disk 
dominates the growth of the SMBH as $M_{\rm sup}$ increases. 
As a result, all the gas accreted from hosts cannot accrete onto 
a central SMBH, but turns into stars in the late phase of mass-supply. 
The above trend does not change for $t_{\rm sup}=10^{7}-10^{9}\,{\rm yr}$ and 
also for changing the star formation efficiency, $C_{*}$.

The averaged mass-supply rate, $\dot{M}_{\rm sup}$, 
is expected to be typically $\approx 0.1-1\, M_{\odot}\,{\rm yr}^{-1}$, 
driven by galaxy mergers (e.g., Mihos \& Hernquist 1996; 
Saitoh \& Wada 2004) or by 
the radiation drag associated with the starbursts 
in hosts (e.g., Umemura 2001; Kawakatu \& Umemura 2002). 
Thus, it may be hard to accrete the gas mass more than 
$\approx 10^{9}M_{\odot}$ from the host galaxy for 
$10^{8}-10^{9}\,{\rm yr}$. 
Even if the gas supply with $1\,M_{\odot}\,{\rm yr}^{-1}$ 
lasts  for $10^{9}\,{\rm yr}$, 
a SMBH does not evolve more than $10^{8}M_{\odot}$. 
This would imply that it is difficult for the only gas accretion 
process due to the turbulent viscosity driven by SN explosions 
to form a SMBH $>\,10^{9}M_{\odot}$ observed 
in the luminous high-$z$ QSOs (e.g., McLure et al. 2006). 
In other words, the direct gas accretion from a host galaxy into 
the accretion disk (sub-pc scale) and/or the mergers of compact 
objects for the formation of SMBH with $>\,10^{8}M_{\odot}$. 

\vspace{5mm}
\epsfxsize=8cm 
\epsfbox{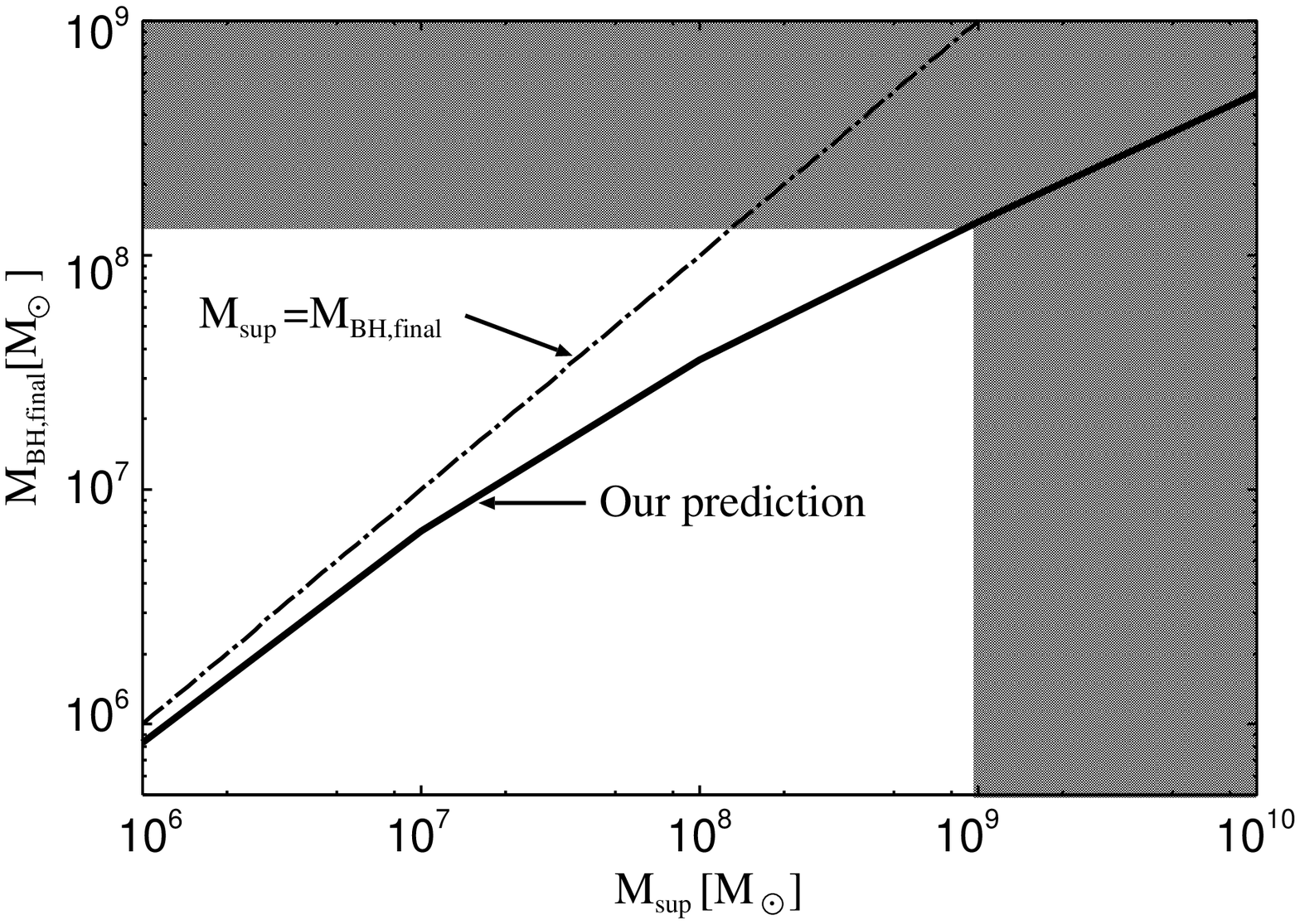}
\figcaption
{
The final SMBH mass, $M_{\rm BH, final}$ against 
the different total accreted mass from host galaxies, $M_{\rm sup}\equiv 
\dot{M}_{\rm sup}t_{\rm sup}$ for $t_{\rm sup}=10^{8}\,{\rm yr}$. 
The thick solid line shows our prediction. 
The dot-dashed line denotes $M_{\rm BH, final}=M_{\rm sup}$. 
We show the parameter space of $M_{\rm sup}$ and $M_{\rm BH, final}$ 
which are hard to achieve by any plausible mass-supply process 
($M_{\rm sup} > 10^{9}M_{\odot}$ and $M_{\rm BH,final} > 10^{8}M_{\odot}$), 
as the shaded region. 
Note that our prediction does not depend on $C_{*}$ and $t_{\rm sup}$ 
(see $\S 3.2$ and $3.3$).
}
\vspace{2mm}

\subsection{Evolution of AGN luminosity and nuclear starburst luminosity}
We examine the evolution of the AGN luminosity and the nuclear-starburst 
luminosity (=the luminosity originating in star formation 
in the circumnuclear disk), 
in order to understand which objects correspond to these two phases 
observationally. 
In this paper, we define these luminosities as follows: 
The AGN luminosity (=the accretion disk luminosity) can be obtained as a function of $\dot{m}_{\rm BH}\equiv \dot{M}_{\rm BH}/\dot{M}_{\rm Edd}$ (Watarai et al. 2000), by assuming two types of accretion disk, namely 
the slim disk (Abramowicz et al. 1988) and 
the standard disk (Shakura \& Sunyaev 1973).

\begin{equation}
L_{\rm AGN}(t)=\left \{
 \begin{array}{l}
 2\left(1+\ln{\frac{\dot{m}_{\rm BH}(t)}{20}}\right)
L_{\rm Edd}(t)\,\,\, ;\dot{m}_{\rm BH}(t) \geq 20, \\ \\
 \left(\frac{\dot{m}_{\rm BH}(t)}{10}\right)
L_{\rm Edd}(t) \,\,\,\,\, ;\dot{m}_{\rm BH}(t) < 20.
 \end{array}\right .
\end{equation}
The nuclear starburst luminosity $L_{\rm Nuc, SB}(t)$ can be given by 
\begin{equation}
L_{\rm Nuc, SB}(t)=0.14\epsilon \dot{M}_{*}(t)c^{2}, 
\end{equation}
where $\epsilon=0.007$ which is the energy conversion efficiency of 
nuclear fusion from hydrogen to helium. 

Figure 6 shows the evolution of AGN luminosity $L_{\rm AGN}(t)$ 
and nuclear-starburst $L_{\rm Nuc, SB}(t)$ 
for $\dot{M}_{\rm sup}=1\, M_{\odot}\, {\rm yr}^{-1}$ 
and $t_{\rm sup}=10^{8}\,{\rm yr}$. 
The mass accretion in units of Eddington mass accretion rate 
$\dot{m}_{\rm BH}$ is larger than 20 until $t=t_{\rm sup}$ 
(see Fig. 2). 
Then, $L_{\rm AGN}(t)$ depends on the evolution of 
$L_{\rm Edd}(t)\propto M_{\rm BH}(t)$ from eq. (15), 
and thus $L_{\rm AGN}(t)$ increases with time. 
During this phase ($t < t_{\rm sup}$), 
the nuclear-starburst luminosity, 
$L_{\rm Nuc, SB}(t) \propto \dot{M}_{*}(t)
\propto M_{\rm g}(t)$, also increases 
with time because $M_{\rm g}(t)$ increases (see Fig. 3). 
%
After $t=t_{\sup}$, both $L_{\rm AGN}(t)$ and $L_{\rm Nuc, SB}(t)$ 
decrease since $M_{\rm g}$ starts to decrease monotonically 
(see Fig. 3).
The AGN luminosity decreases drastically after $t=t_{\rm crit}$, 
reflecting the phase change of accretion (see Fig. 2). 
After $t=t_{\rm stop}$ 
$L_{\rm Nuc, SB}$ becomes zero because the circumnuclear disk 
is gravitationally stable as mentioned in $\S 3.1$. 
In summary, there are four phases in $L_{\rm AGN}(t)$ and 
$L_{\rm Nuc, SB}(t)$; 
(i) the super-Eddington luminosity phase: $L_{\rm AGN} > L_{\rm Edd}$ 
($t < t_{\rm sup}$), 
(ii) the sub-Eddington luminosity phases: 
$L_{\rm AGN}=(0.1-1)L_{\rm Edd}$ ($t_{\rm sup} < t < t_{\rm crit}$),
(iii) the starburst-luminosity dominated phase: 
$L_{\rm AGN} < L_{\rm Nuc, SB} \ll L_{\rm Edd}$ 
($t_{\rm crit} < t < t_{\rm stop}$), 
and (iv) the post AGN/starburst phase: $L_{\rm AGN} \ll L_{\rm Edd}$ \& 
$L_{\rm Nuc, SB}=0$ ($t > t_{\rm stop}$). 

\vspace{5mm}
\epsfxsize=8cm 
\epsfbox{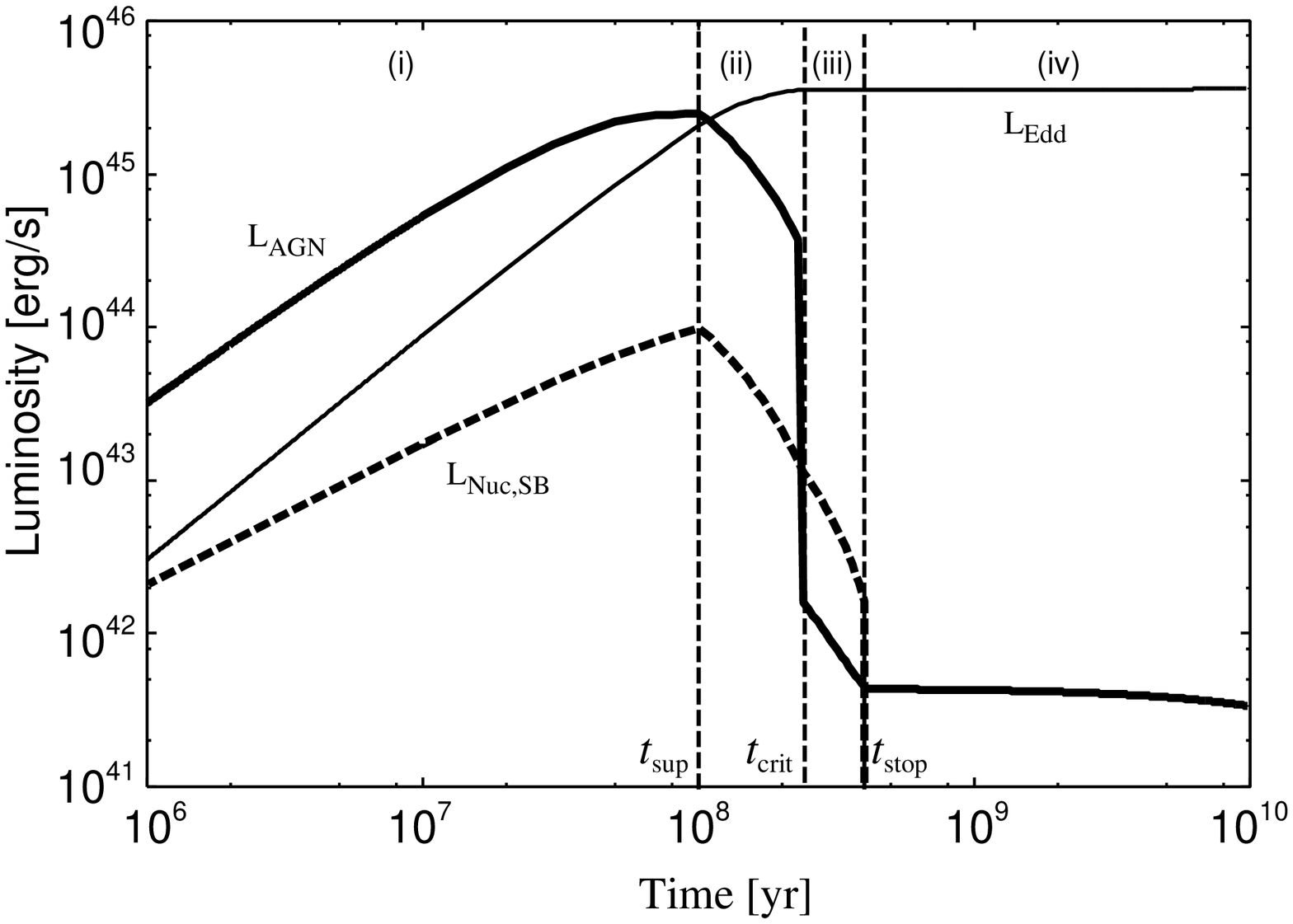}
\figcaption
{
Time evolution of the AGN luminosity, $L_{\rm AGN}(t)$, 
that of the nuclear starburst luminosity, $L_{\rm Nuc, SB}(t)$ 
and that of the Eddington luminosity, $L_{\rm Edd}(t)$. 
We can divide this into the four phases as follows: 
(i) $t < t_{\rm sup}$, 
(ii) $t_{\rm sup} < t < t_{\rm crit}$, 
(iii) $t_{\rm crit} < t < t_{\rm stop}$ and 
(iv) $t > t_{\rm stop}$. 
These phases (i), (ii), (iii) and (iv) correspond to 
the super-Eddington luminosity phase, the sub-Eddington luminosity 
phase, the starburst-luminosity dominated phases and 
the post-AGN/Starburst phase, respectively.
}
\vspace{2mm}

Phase (i) might be a good model for 
narrow-line type I Seyfert galaxies (NLS1s) 
because NLS1s are super-Eddington AGNs (e.g., Mathur et al. 2001; 
Kawaguchi 2003). 
On the other hands, if the mass supply is 
extremely high (e.g., $\dot{M}_{\rm sup}=10M_{\odot}\,{\rm yr}^{-1}$) 
and consequently the circumnuclear region 
is covered by dense gas, then this phase might be observed as 
ULIRGs. Note that some kind of ULIRGs show properties similar to NLS1s 
(e.g., Teng et al. 2005; Hao et al. 2005; Kawakatu et al. 2007a). 
If this is the case, we predict that {\it a gas-rich massive torus} 
($m_{\rm disk} > 1$ and $f_{\rm g}=1$) exists at their center. 
Such a massive torus can be observed through carbon monoxide and 
hydrogen cyanide molecular emission using the 
Atacama Large Millimeter Array instrument (ALMA) for 
NLS1s and ULIRGs (e.g., Wada \& Tomisaka 2005; Kawakatu et al. 2007b; Yamada, 
Wada \& Tomisaka 2007). 
Judging from the luminosity ratio ($L_{\rm AGN}/L_{\rm Edd}=0.1-1$), 
in phase (ii) the AGN is still bright and thus this might be BLS1s or QSOs. 
In this phase, the relatively small $m_{\rm disk}$ and $f_{\rm g}$ are 
predicted, i.e., $m_{\rm disk}=0.2-1$ and $f_{\rm g}\approx 0.1-0.5$.
After phase (ii), the AGN luminosity becomes smaller by three order of 
magnitude than its peak luminosity. 
Thus, phases (iii) and (iv) may correspond to the low luminosity AGNs (LLAGNs) 
associated with and without the nuclear starburst, respectively. 
As seen in Fig. 3, {\it a geometrically thin 
massive stellar disk}, i.e., $M_{*} > M_{\rm BH}$ 
and $f_{\rm g}\approx 10^{-2}$ is expected in this post-AGN phase. 
Thus, this indicates that the LLAGNs do not have obscuring tori 
envisaged in the AGN unified model because no stars form near the 
inner radius of the circumnuclear disk. 
In other words, we predict the absence of obscuring tori in LLAGNs. 
This seems to be consistent with observations that 
most low luminosity radio galaxies (FR I) show no evidence of dusty tori 
expected in normal AGNs (e.g., Chiaberge et al.1999). 
As an alternative explanation, Elizur \& Shlosman (2006) 
suggested the torus disappears when the AGN luminosity decreases 
below $\sim 10^{42}\,{\rm erg}\,{\rm s}^{-1}$ because the mass accretion 
in the accretion disk cannot maintain the required cloud outflow
(see also H\"{o}nig \& Beckert 2007).
In order to understand the origin of absence of dusty tori 
in LLAGNs, it is essential to explore the spatial distribution of 
young stars in them.
In addition, 
since such a stellar disk is a remnant of a gas-rich massive torus 
in the early phase of SMBH growth, its discovery would strongly support 
the idea that LLAGNs are dead Seyfert galaxies or QSOs. 
Therefore, the observations of the physical states of a circumnuclear 
disk on $\leq 100$ pc, e.g., $m_{\rm disk}$ and $f_{\rm g}$, 
for different type of AGNs would be useful to understand AGN evolution. 

Finally, we summarize a coevolution scenario of SMBHs and 
circumnuclear disks in Fig. 7. 
According to our knowledge obtained in this paper, 
we might guess that the final state of AGNs depends on the 
mass supply rate $\dot{M}_{\rm sup}$ and the duration of the mass-supply 
$t_{\rm sup}$. 
In details, we will discuss this issue in our on-coming paper. 

\vspace{5mm}
\epsfxsize=8cm 
\epsfbox{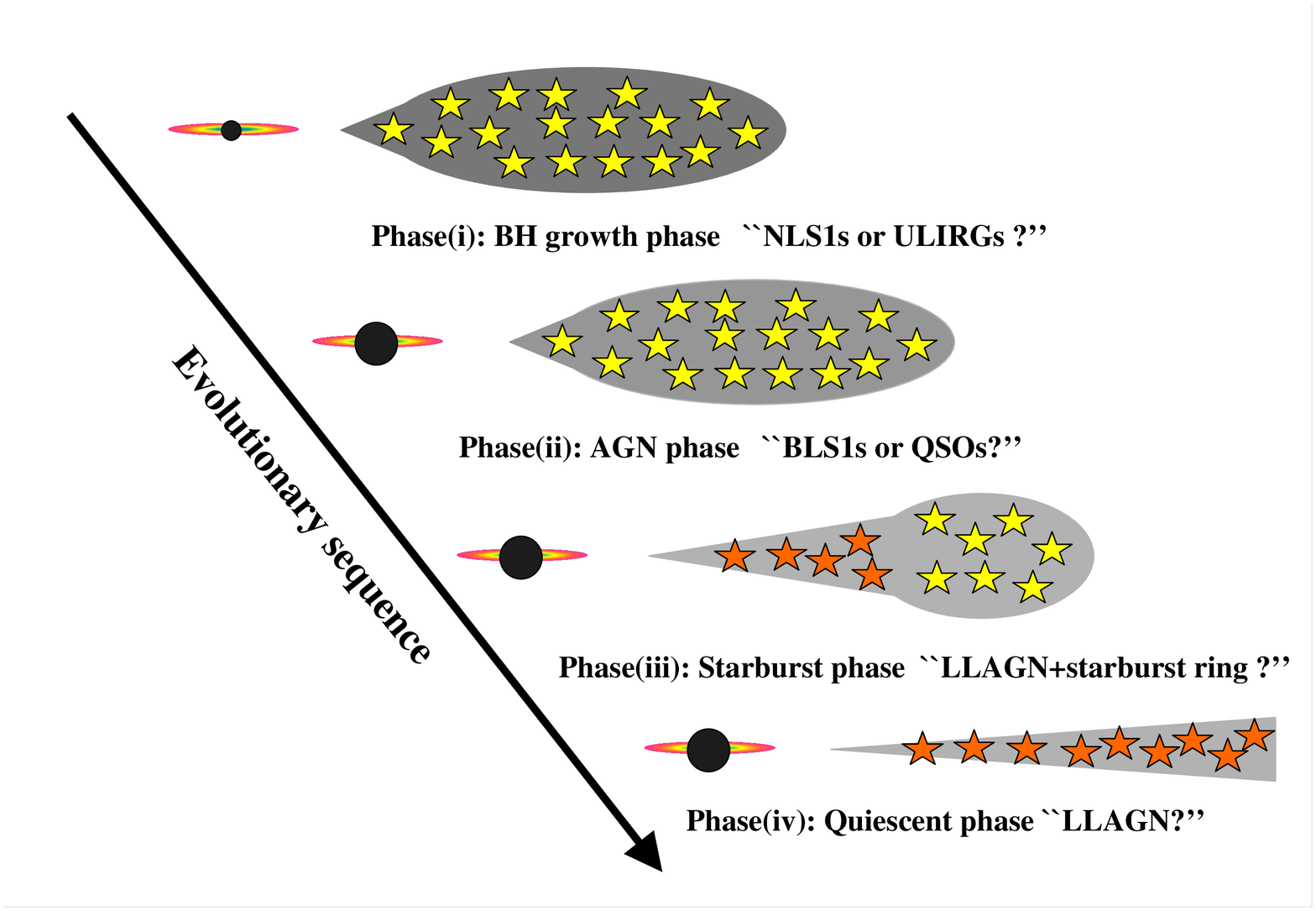}
\figcaption
{
Schematic sketch for the coevolution of a SMBH and a 
circumnuclear disk. 
Star symbols (yellow) represent young stars. 
Star symbols (red) denote old stars. 
The darkness of gray color shows the gas fraction ($f_{\rm g}$), 
that is, the darker gray represents higher gas fraction. 
}
\vspace{2mm}

\subsection{AGN luminosity vs. Nuclear-starburst luminosity relation}
Based on our model, we present the correlation between AGN luminosity 
and nuclear-starburst luminosity. 
In Fig. 8, we plot $L_{\rm Nuc, SB}$ against $L_{\rm AGN}$ 
for the different $M_{\rm BH, final}$ (or $M_{\rm sup}$). 
The higher concentration of data points represents that 
the evolution is slow. 
We find that AGN luminosity positively correlates with 
nuclear-starburst luminosity for bright AGNs (right branches of Fig. 8), 
which corresponds to the high-accretion phase, 
that is, phase (i) and phase (ii) defined in $\S 3.4$. 
This trend is consistent with the observational results for nearby 
Seyfert galaxies (Imanishi \& Wada 2004; Watabe et al. 2008). 
Moreover, $L_{\rm Nuc, SB}/L_{\rm AGN}$ becomes larger as 
$L_{\rm AGN}$ increases.
This can be explained as follows: 
$L_{\rm Nuc, SB}/L_{\rm AGN}\,(\propto \dot{M}_{*}/\dot{M}_{\rm BH})$ 
is larger as $M_{\rm g}$ increases because of $\dot{M}_{*}/\dot{M}_{\rm BH}
\propto M_{\rm g}^{1/2}$ (see $\S 3.1$). 
Considering $L_{\rm AGN}\propto M_{\rm BH}\propto M_{\rm g}^{1/2}$ 
during $t < t_{\rm sup}$, 
$L_{\rm Nuc, SB}/L_{\rm AGN}$ increases with $L_{\rm AGN}$. 
We should mention that it is worth examing this trend by future observations.

On the other hand, the $L_{\rm AGN}$ vs. $L_{\rm Nuc, SB}$ 
relation in LLAGNs (left branches) does not follow that in bright AGNs. 
In this phase, the range of AGN luminosity is independent of $L_{\rm Nuc, SB}$, i.e., $L_{\rm AGN}=10^{41}-10^{42}\,{\rm erg}\,{\rm s}^{-1}$ for 
$L_{\rm Nuc, SB}=10^{40}-10^{44}\,{\rm erg}\,{\rm s}^{-1}$, 
because $\dot{M}_{\rm BH}$ is basically determined by the sound velocity 
in the disk. 
Since the nuclear-starburst luminosity, 
$L_{\rm Nuc, SB}\propto \dot{M}_{*}\propto M_{\rm g}$ 
increases with $M_{\rm g}$, 
no tight correlation between $L_{\rm Nuc, SB}$ and 
$L_{\rm AGN}$ is seen in LLAGNs. 
In summary, we suggest that the AGN-starburst connection depends on the 
evolution of the AGN activity. 
In order to test our model, it is essential to compare 
the AGN-starburst luminosity relation for bright AGNs 
(e.g., Seyfert galaxies and QSOs) with that for LLAGNs. 

\vspace{5mm}
\epsfxsize=8cm 
\epsfbox{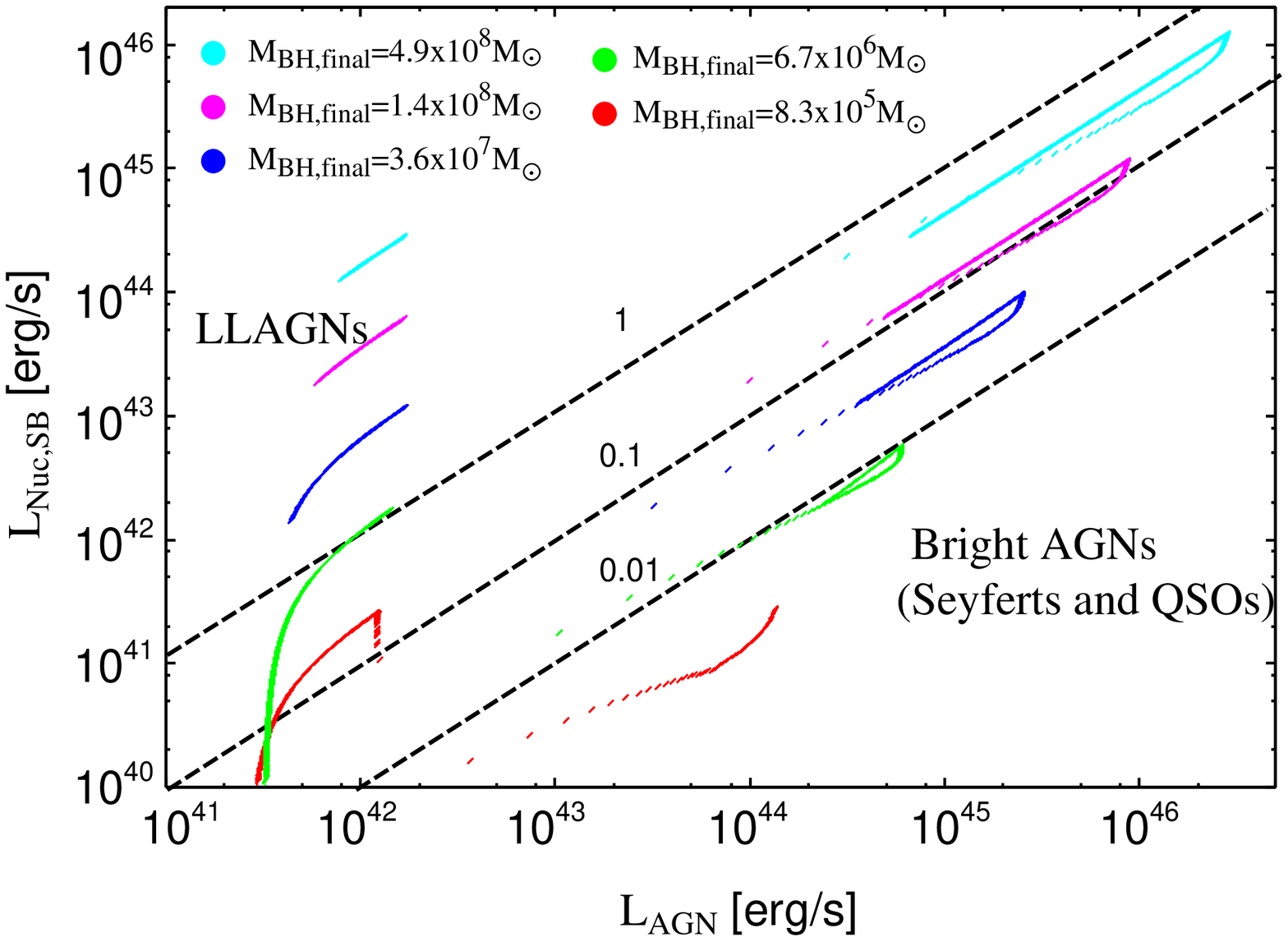}
\figcaption
{
The nuclear starburst luminosity, $L_{\rm Nuc, SB}$ against 
the AGN luminosity, $L_{\rm AGN}$. 
The dots with different colors denote the different final BH mass, 
$M_{\rm BH, final}$. 
The dashed lines represent the luminosity ratio, 
$L_{\rm Nuc, SB}/L_{\rm AGN}=0.01, \, 0.1\,{\rm and}\, 1$, respectively.
}
\vspace{2mm}

\section{Discussion}
\subsection{Effects of AGN outflow}
Through this paper, we assume all the gas is accreted by a 
central BH, i.e.,  $\dot{M}_{\rm BH}=\dot{M}(r_{\rm in})$ 
even when the mass accretion rate at $r_{\rm in}$ 
exceeds the Eddington limit (We call this model ``{\it model (a)}'').
According to the recent two dimensional radiation hydrodynamic 
simulations (e.g., Ohsuga et al. 2005), the super-Eddington 
mass accretion actually can be achieved, but they mentioned 
that the large part of accreting gas turns into  outflow because of the 
strong radiation pressure from the accretion disk 
(Ohsuga et al. 2005; Ohsuga 2007). 

We here assume that the dynamics of circumnuclear disk is not 
affected by the outflow. 
The two models of AGN outflow are possible when  
$\dot{M}(r_{\rm in}) > \dot{M}_{\rm Edd}$ as follows;

(i) The BH grows by a super Eddington rate but the 
part of accreting matter forms outflow, ``{\it model (b)}''. 

\begin{eqnarray}
&&\dot{M}_{\rm outflow}=f_{\rm w}\dot{M}(r_{\rm in})\,\,{\rm and} \nonumber  \\
&&\dot{M}_{\rm BH}=(1-f_{\rm w})\dot{M}(r_{\rm in}) \nonumber, 
\end{eqnarray}
where we assume $f_{\rm w}=0.1$ from the results of Ohsuga (2007). 

(ii) The growth of BHs is regulated by the 
Eddington limit and the remnant of accreting gas turns into the outflow, 
``{\it model (c)}''. 
\begin{eqnarray}
&&\dot{M}_{\rm outflow}=\dot{M}(r_{\rm in})-\dot{M}_{\rm BH}\,\,{\rm and}\nonumber  \\
&&\dot{M}_{\rm BH}=\epsilon_{\rm BH}^{-1}\dot{M}_{\rm Edd} \nonumber,
\end{eqnarray}
where $\epsilon_{\rm BH}=0.05$ is the energy conversion efficiency. 

Figure 9 is the same as Fig. 2 and Fig. 3, but the outflow rate 
($\dot{M}_{\rm outflow}$) and the total outflow mass ($M_{\rm outflow}$) are 
considered based on {\it model (b)}, 
where $M_{\rm outflow}\equiv \int_{0}^{t}\dot{M}_{\rm outflow
}(t^{\prime})dt^{\prime}$.
During $t < t_{\rm of}$ the SMBH grows by the super-Eddington mass accretion 
rate, but most of accreting mass from the gas disk is blown away 
as the AGN outflows, where $t_{\rm of}$ is the time when 
$\dot{M}(r_{\rm in})=\epsilon_{\rm BH}^{-1}\dot{M}_{\rm Edd}$. 
After $t=t_{\rm of}$, all the gas accrete onto a central BH by less 
than the Eddington limit (i.e., no AGN outflow). 
After $t=t_{\rm crit}$ the evolutions are the same as the case without 
the outflow, i.e., {\it model (a)}.
As seen in Fig. 9, we find that the final mass of SMBHs is 
$M_{\rm BH, final}\simeq 8\times 10^{6}M_{\odot}$, 
which is a factor four smaller compared with the prediction by {\it model (a)}. 
\vspace{5mm}
\epsfxsize=8cm 
\epsfbox{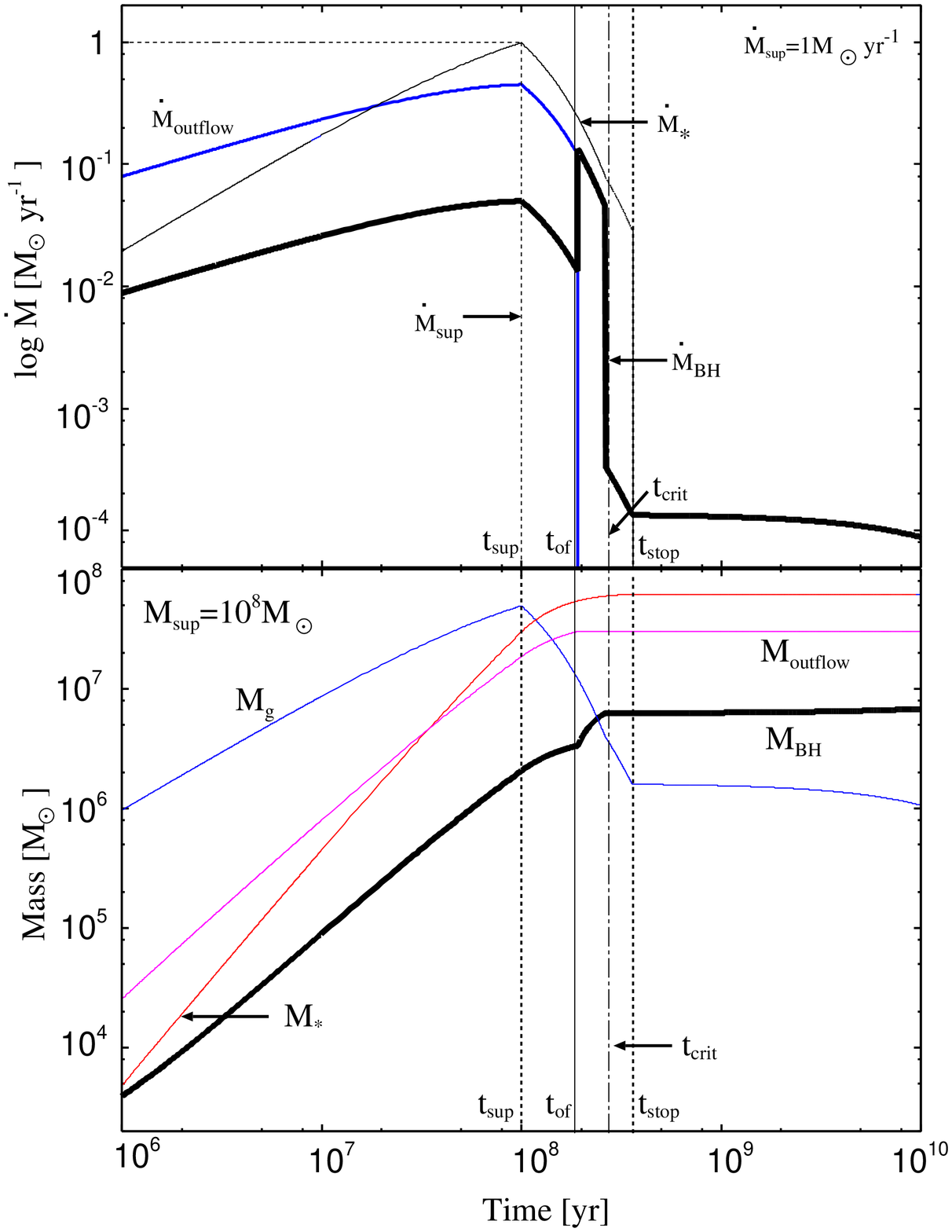}
\figcaption
{
Same as Fig. 2 and Fig. 3, but the outflow rate ($\dot{M}_{\rm outflow}$) 
and the total outflow mass ($M_{\rm outflow}$) are newly added for the 
{\it model (b)}, i.e., super-Eddington growth of SMBHs with the AGN outflow. 
Here $t_{\rm of}$ is the time when $\dot{M}(r_{\rm in})=
\epsilon^{-1}_{\rm BH}\dot{M}_{\rm Edd}$.
}
\vspace{2mm}
In the {\it model (a)}, the super-Eddington accretion mainly determines 
the final SMBH mass (see Fig. 2 and Fig. 3). 
But, in {\it model (b)} the contribution of super-Eddington accretion is 
comparable to that of Eddington accretion for the final SMBH mass. 
This indicates that the contribution of super-Eddington accretion 
depends on the strength of AGN outflows, i.e., $f_{\rm w}$. 
Therefore, it will be worth revealing which physics controls 
$f_{\rm w}$ by using the radiation hydrodynamic simulations. 

On the other hands, we show the case of Eddington limited BH growth 
(i.e., {\it model (c)}) in Fig. 10. 
Apparently, the BH growth is greatly suppressed by AGN outflow and 
consequently the final SMBH is order of magnitude less than 
{\it model (a)}, namely $M_{\rm BH, final}\simeq 10^{6}M_{\odot}$. 
In summary, we find that the evolution of SMBHs are closely 
linked with not only the state of circumnuclear disk, but also 
the strength of AGN outflow and the conditions under 
which the super-Eddington growth occurs. 

\vspace{5mm}
\epsfxsize=8cm 
\epsfbox{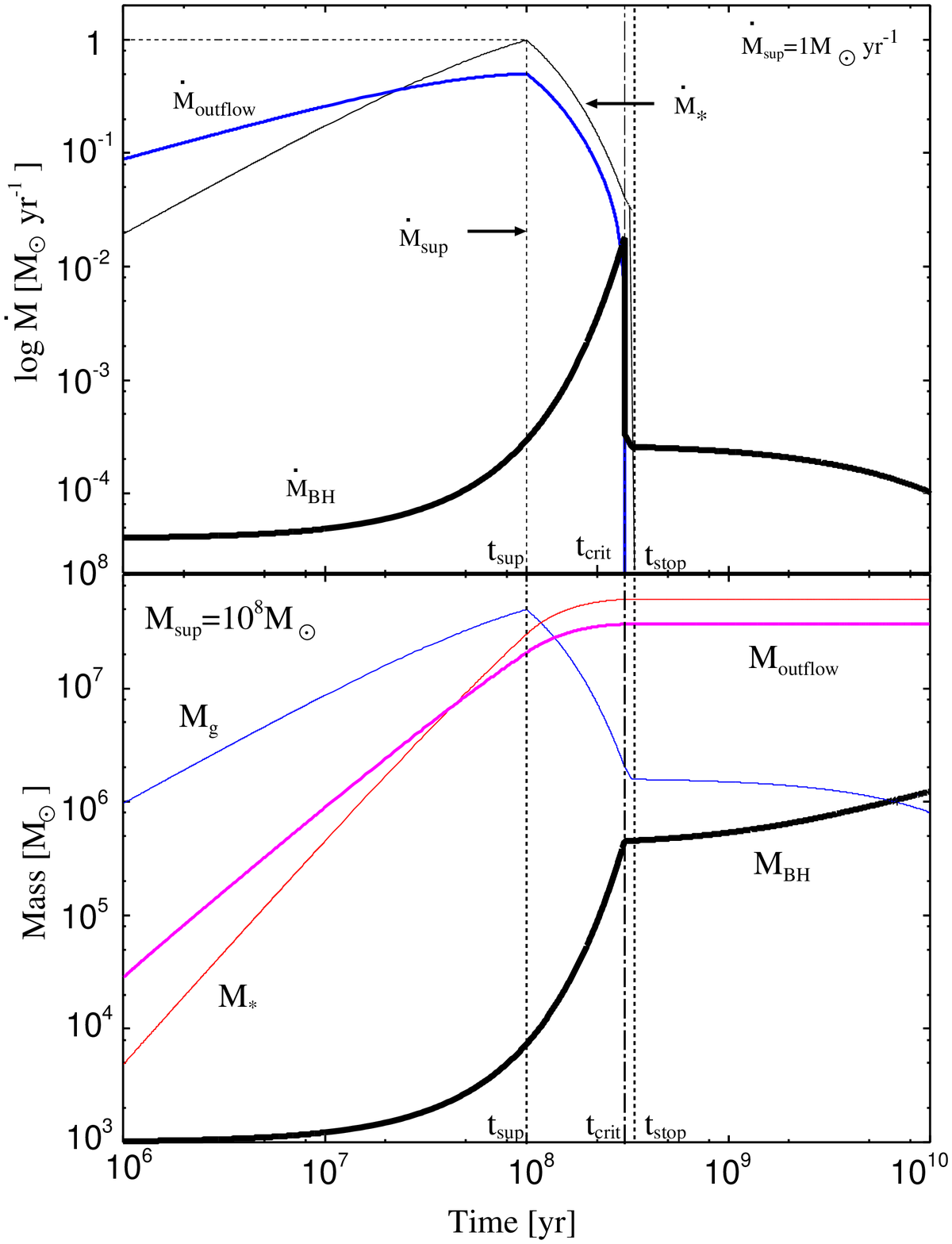}
\figcaption
{
Same as Fig. 2 and Fig. 3, but the outflow rate ($\dot{M}_{\rm outflow}$) 
and the total outflow mass ($M_{\rm outflow}$) are newly added for the 
{\it model (c)}, i.e., Eddington limited growth of SMBHs with the AGN outflow.
}
\vspace{2mm}

\subsection{Effects on the UV radiation pressure}
So far, we have ignored the effect of radiation pressure, but 
the UV radiation from young stars in the gas disks may affect on the 
vertical structure of disk. 
Thus, we examine whether the radiation pressure is really negligible 
when the disk geometry is determined by the SN feedback. 
If the disk is optically thick to the UV radiation, 
the radiation pressure ($P_{\rm rad}$) is given by 
\begin{equation}
P_{\rm rad}=l_{\rm SB}/c\simeq 0.14\epsilon S_{*}(r)h_{\rm tub}(r)c, 
\end{equation}
where $h_{\rm tub}(r)$ is determined by eq. (5). 
Here $l_{\rm SB}=0.14\epsilon S_{*}(r)h(r)c^{2}$ is the 
starburst luminosity per unit area.

As a result, the ratio of the radiation pressure and the turbulent pressure 
via SN explosions ($P_{\rm tub}=\rho_{\rm g}v_{\rm t}^{2}$) can be obtained 
as follows:
\begin{equation}
\frac{P_{\rm rad}}{P_{\rm tub}}\sim \eta_{-3}^{-1/2}
\left(\frac{t_{\rm dyn}}{t_{*}}\right)^{1/2}, 
\end{equation}
where $t_{\rm dyn}=(r^{3}/GM_{\rm BH})^{1/2}$ is the dynamical timescale. 
Because of $t_{\rm dyn}(r_{\rm out})/t_{\rm *} \sim 10^{-2}$ for 
the relatively high star formation efficiency, 
i.e., $C_{*}=10^{-8}\,{\rm yr}^{-1}$, the radiation pressure 
is neglected as far as $\eta > 10^{-5}M_{\odot}^{-1}$. 
This condition ($\eta > 10^{-5}M_{\odot}^{-1}$) is satisfied 
for a turbulent pressure supported starburst 
disk modeled by WN02 using three-dimensional hydrodynamical 
simulations. Thus, our conclusions do not change significantly even if 
we consider the UV radiation pressure. 

\subsection{Origin of a bias in AGN formation}
Thanks to a growing number of observations at different redshifts 
in multi-wavelength, it has been suggested that the density of more luminous 
AGNs associated to more massive SMBHs peaks earlier in the history of 
the universe than that of low luminosity ones (Miyaji, et al. 2000; 
Ueda et al. 2003; Hasinger et al. 2005; Bongiorno et al. 2007). 
This evolution has been dubbed as AGN cosmic downsizing. 
It is still unclear why the growth timescale of SMBHs is shorter 
in more massive SMBHs, in other words, why the growth time is highly 
biased in an environment to form massive SMBHs. 

From eq. (14), the final SMBH mass can be described as 
\begin{equation}
M_{\rm BH,final}\approx 10^{8}M_{\odot}
\left(\frac{C_{*}}{10^{-8}\,{\rm yr}^{-1}}\right)^{2}
\left(\frac{t_{\rm growth}}{5\times 10^{8}\,{\rm yr}}\right)^{2}. 
\end{equation} 

The AGN downsizing might be explained by our result, 
if larger $M_{\rm BH, final}$ grows 
in the circumnuclear disk with a higher star formation efficiency. 
For instance, Fig. 11 shows that the growth timescale for $M_{\rm BH,final}=10^{8}M_{\odot}$ is shorter than that for $M_{\rm BH,final}=10^{7}M_{\odot}$, 
i.e., $t_{\rm growh}(M_{\rm BH,final}=10^{8}M_{\odot})\approx 10^{8}\,{\rm yr}
$ and $t_{\rm growh}(M_{\rm BH,final}=10^{7}M_{\odot})\approx 10^{9}\,{\rm yr}$, if $C_{*}(M_{\rm BH,final}=10^{8}M_{\odot})$ is $\approx 30$ times as large 
as $C_{*}(M_{\rm BH,final}=10^{7}M_{\odot})$. 
According to Wada \& Norman (2007), they showed that the star formation rate 
increases extensively as a function of average gas density. 
Since the massive galaxies could be formed in an environment with 
a higher average gas density, the star formation efficiency could be 
$\approx 10$ times larger if the average gas density is $\approx 
10$ times higher. 
As a consequence, the UV radiation field would be stronger in more 
massive galaxies. 
The recent radiation hydrodynamics simulations showed that the 
critical density for the star formation can be affected by the 
UV radiation (e.g., Susa \& Umemura 2004). 
These works indicate that the critical density is larger 
for a stronger UV radiation field. 
In addition, the star formation efficiency, $C_{*}$ would be 
larger for a larger critical density at given average gas 
density (Wada \& Norman 2007). 
Therefore, since the circumnuclear disk in more massive galaxies 
is exposed to the stronger UV radiation, 
the star formation efficiency is larger for the gas disk in 
massive galaxies. 
Therefore, the strength of UV radiation field surrounding the galactic 
nuclei might control the timescale for SMBH growth. 
In order to clarify this issue, it is essential to examine the 
star formation process in the gas disk under an intense UV field 
with the three-dimensional radiation transfer. 

\vspace{5mm}
\epsfxsize=8cm 
\epsfbox{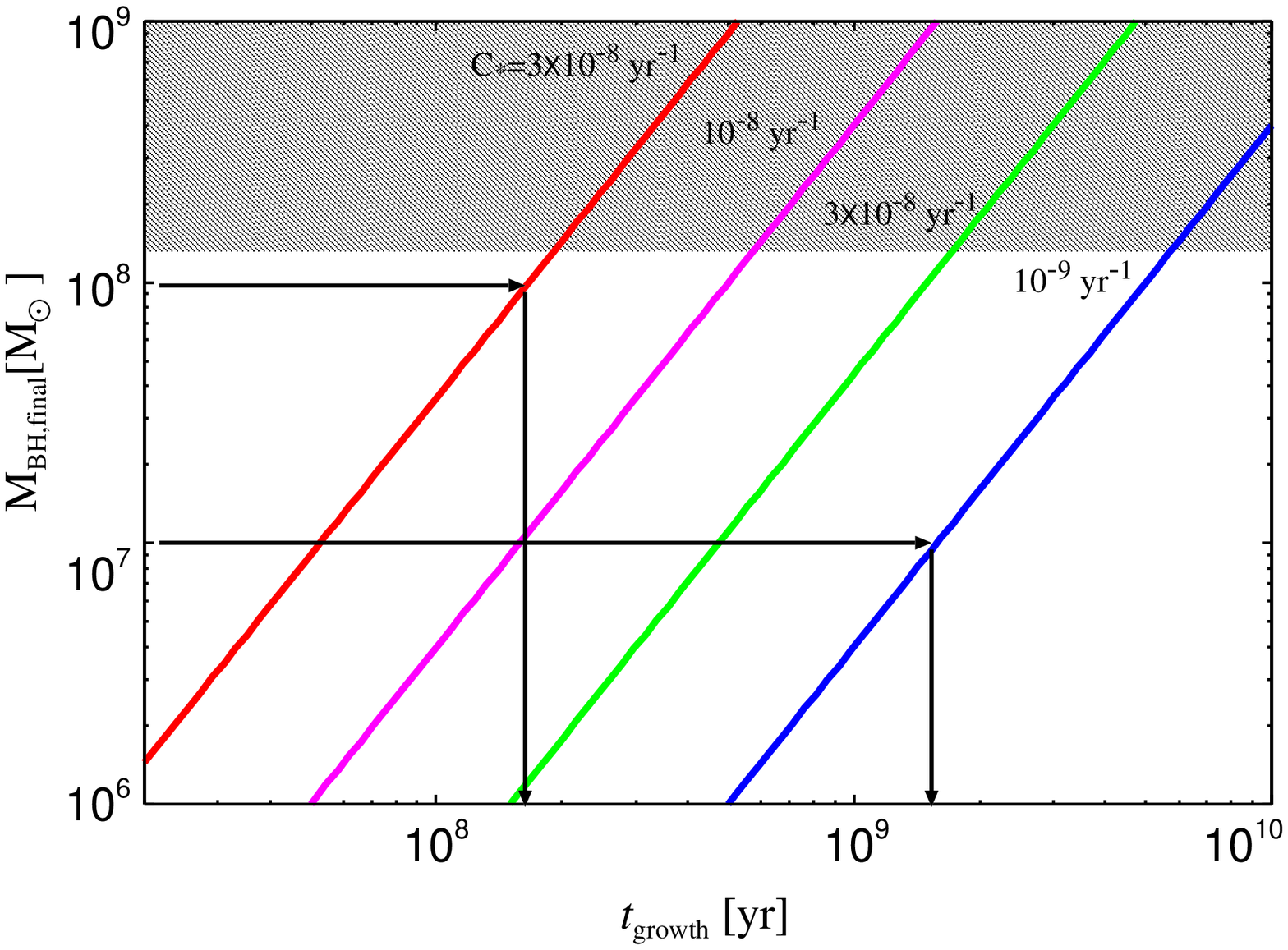}
\figcaption
{
The final SMBH mass, $M_{\rm BH, final}$ against the timescale of 
the SMBH growth without the AGN outflow ({\it model (a)}). 
The solid lines with different colors correspond to 
the different star formation efficiency. 
The shaded region denotes the parameter space of $M_{\rm BH, final}$ 
which is difficult to achieve in the present model.
}
\vspace{2mm}

\section{Conclusions} 

We have elucidated the physical role of the circumnuclear disk 
in relation to SMBH growth.  
To this end, we have constructed a new model of SMBH growth, 
taking into account the evolution of physical states of the circumnuclear disk 
formed by the mass-supply from the host galaxy. 
In the present model, we consider the two regimes of gas accretion 
depending on the gravitational stability of the disk. 
Our main conclusions are summarized as follows:

\begin{enumerate}
\item 
Not all the gas in the disk accrete onto the SMBH, i.e., 
$\dot{M}_{\rm BH}=0.1-0.3M_{\odot}\,{\rm yr}^{-1}$ 
for $\dot{M}_{\rm sup}=1\, M_{\odot}\,{\rm yr}^{-1}$ 
because the part of gas is used to form stars in the disk. 
This high accretion phase changes to the low accretion phase 
with $\dot{M}_{\rm BH}\sim 10^{-4}M_{\odot}\,{\rm yr}$. 
But the transition takes $\sim 10^{8}\,{\rm yr}$ 
after the mass supply from hosts is stopped. 
This timescale is basically determined by the star formation 
timescale, $t_{*}\simeq C_{*}^{-1}\sim 3\times 10^{7}\,{\rm yr}$. 
Through this evolution, the gas rich disk turns into the gas poor 
stellar disk. In the high accretion phase, a super Eddington accretion 
is possible, and thus the existence of gas rich circumnuclear disks 
($M_{\rm g} > M_{\rm BH}$) is a condition which the super Eddington accretion 
onto a central SMBH keeps.
In the present model, 
the two phases of SMBH growth depend on whether stars can form 
in the inner region of the circumnuclear disk. 
This would imply that the AGN activity depends on 
the spatial distribution of young stars in the circumnuclear disk. 
In order to examine this, infrared observations with 
high spatial resolution 
(e.g. Very Large Telescope Interferometer (VLTI), Space Infrared Telescope for Cosmology \& Astrophysics (SPICA) and 3m Atlantic infrared telescope) are 
crucial, because young stars would be buried in the optically thick 
circumnuclear disk.

\item
Evolution of nuclear activities are divided into four phases;
(i) the super-Eddington luminosity phase, (ii) the 
sub-Eddington luminosity phase, (iii) the starburst-luminosity 
dominated phases, and (iv) the post-AGN phase. 
Judging from the luminosity ratio, $L_{\rm AGN}/L_{\rm Edd}$, 
NLS1s or ULIRGs may correspond to phase (i). If this is the case, 
it is predicted that they have gas-rich tori, 
whose masses are larger than the mass of their central SMBHs. 
It indicates that the SMBH growth proceeds 
in the phase surrounded by gas-rich tori.
Phase (ii) may correspond to BLS1s or QSOs. 
In this phase, we predict that 
the gas fraction of the circumnuclear disk is smaller than that 
for NLS1s or ULIRGs, i.e., $f_{\rm g}\sim 10-50\%$.
Finally, phases (iii) and (iv) may correspond 
to the low-luminosity AGNs (LLAGNs). 
According to our prediction, LLAGNs have gas-poor (gas fraction 
$f_{\rm g}\sim 1\%$) circumnuclear disks. 
This result suggests that the physical states of circumnuclear 
gas or stellar systems are important to understand the nature 
of different types of AGNs for gas components 
with ALMA and/or VLTI for stellar components. 

\item
The final SMBH mass ($M_{\rm BH, final}$) is not 
proportional to the total gas mass supplied from the host galaxy 
($M_{\rm sup}$) during the hierarchical formation of 
galaxies, i.e., $M_{\rm BH, final}/M_{\rm sup}$ decreases 
with $M_{\rm sup}$, because the star-formation rate overcomes 
the growth rate of SMBHs as $M_{\rm sup}$ increases. 
Considering the plausible mass accretion processes from the 
host galaxy, our results may indicate that it is difficult to form 
$\approx 10^{9}M_{\odot}$ SMBHs observed at 
high-$z$ QSOs, only by the gas accretion process due to the turbulent 
viscosity via SN explosions. 
This might imply the importance of alternative mechanisms for the formation 
of $\sim 10^{9}M_{\odot}$ SMBH. 

\item
The AGN luminosity ($L_{\rm AGN}$) 
tightly correlates with the nuclear-starburst luminosity ($L_{\rm Nuc, SB}$) 
in the bright AGNs (Seyfert galaxies and QSOs). 
We also predict that $L_{\rm Nuc, SB}/L_{\rm AGN}$ is larger 
for more luminous AGNs (i.e., QSOs). 
This is not the case in the LLAGNs. 
Therefore, the preset model predicts 
the $L_{\rm AGN}$ vs. $L_{\rm Nuc, SB}$ relation depends on 
the evolution of AGN activity.

\item
The AGN outflow from the circumnuclear disk greatly 
suppresses the growth of the SMBHs, especially if the 
BH growth is limited by the Eddington rate 
when the mass accretion from the disk exceeds the Eddington limit. 
This indicates that the strength of AGN outflows affects 
on the evolution of SMBHs and final mass of SMBHs. 

\end{enumerate}

\acknowledgments 
We appreciate the detailed reading the fruitful comments of 
anonymous referees. 
We thank K. Ohsuga and R. I. Davies and T. Saitoh for useful discussions. 
K. W. is supported by Grant-in-Aids for Scientific Research 
(15684003 and 16204012 [K. W.]) of JSPS.


\end{document}